\documentclass[11pt,prd,aps,a4paper]{revtex4}

\usepackage{graphicx}
\usepackage{amssymb}
\usepackage{epstopdf}

\newcommand{\be}{\begin{equation}}
\newcommand{\ee}{\end{equation}}
\newcommand{\bea}{\begin{eqnarray}}
\newcommand{\eea}{\end{eqnarray}}

\begin{document}


\title{Extraction of HQE parameters from unquenched lattice data\\[2mm] on pseudoscalar and vector heavy-light meson masses \vspace{1cm}}

\author{P.~Gambino} \affiliation{Dipartimento di Fisica, Universit\`a di Torino and INFN Sezione di Torino,\\ Via P. Giuria 1, I-10125 Torino, Italy}
\author{A.~Melis} \affiliation{Dep.~de F\'{\i}sica Te\`{o}rica and IFIC, Universitat de Val\`{e}ncia, Dr.~Moliner 50, E-46100 Burjassot, Spain}
\author{S.~Simula} \affiliation{Istituto Nazionale di Fisica Nucleare, Sezione di Roma Tre,\\ Via della Vasca Navale 84, I-00146 Rome, Italy}

\begin{abstract}
\vspace{1.0cm}
We present a precise lattice computation of pseudoscalar and vector heavy-light meson masses for heavy-quark masses ranging from the physical charm mass up to $\simeq 4$ times the physical b-quark mass. We employ the gauge configurations generated by the European Twisted Mass Collaboration (ETMC) with $N_f = 2+1+1$ dynamical quarks at three values of the lattice spacing ($a \simeq 0.062, 0.082, 0.089$ fm) with pion masses in the range $M_\pi \simeq 210 - 450$ MeV. The heavy-quark mass is simulated directly on the lattice up to $\simeq 3$ times the physical charm mass. The interpolation to the physical $b$-quark mass is performed using the ETMC ratio method, based on ratios of the meson masses computed at nearby heavy-quark masses, and adopting the kinetic mass scheme. The extrapolation to the physical pion mass and to the continuum limit yields $m_b^{\rm kin}(1~\mbox{GeV}) = 4.61 (20)$ GeV, which corresponds to $\overline{m}_b(\overline{m}_b) = 4.26 (18)$ GeV in the $\overline{MS}$ scheme. The lattice data are analyzed in terms of the Heavy Quark Expansion (HQE) and the matrix elements of dimension-4 and dimension-5 operators are extracted with a good precision, namely: $\overline{\Lambda} = 0.552 (26)$ GeV, $\mu_\pi^2 = 0.321 (32)$ GeV$^2$ and $\mu_G^2(m_b) = 0.253 (25)$ GeV$^2$.
The data also allow for a rough estimate of the dimension-6 operator matrix elements.
As the HQE parameters play a crucial role in the inclusive determination of the Cabibbo-Kobayashi-Maskawa matrix elements $V_{ub}$ and $V_{cb}$,  their precise determination on the lattice may eventually validate and improve the analyses based on fits to the semileptonic moments.

\end{abstract}

\maketitle

\newpage

\section{Introduction}
\label{sec:intro}

The precise determination of the Cabibbo-Kobayashi-Maskawa (CKM) matrix element $V_{cb}$ is crucial for testing the Standard Model (SM) predictions for the rare decays driven by the charged current $b \to c$ transition and in the quest for new physics effects.
The information on the CKM entry $V_{cb}$ can be obtained from both inclusive and exclusive semileptonic $B$-meson decays.
In the first case the Operator Product Expansion (OPE) is usually adopted to describe the non-perturbative hadronic physics in terms of few parameters that can be extracted from experimental data on inclusive $B \to X_c \ell \nu_\ell$ decays together with the CKM element $V_{cb}$ (see, e.g., Ref.~\cite{Gambino:2016jkc,Alberti:2014yda} and therein).
In the second case the relevant hadronic inputs are the semileptonic form factors describing the $B \to D^* (D) \ell \nu_\ell$ decays.
The latter are computed using non-perturbative methods, like lattice QCD (LQCD) simulations.
As is well known, there is a long-standing tension of about 3 standard deviations between the values of $V_{cb}$ obtained from inclusive or exclusive semileptonic $B$-meson decays \cite{PDG}, although new evidence suggests that part of this discrepancy may be due to 
the way the experimental data have been analysed \cite{Bigi:2017njr}.

The aim of this work is to address the lattice determination of some of the parameters appearing in the OPE analysis of the inclusive $B$-meson decays.
Indeed, the same parameters  (or combinations thereof) also appear as coefficients of the Heavy Quark Expansion (HQE) for the pseudoscalar (PS) and vector (V) heavy-light meson masses. 
So far, only the charmed and beauty mesons masses, $M_{D^{(*)}}$ and $M_{B^{(*)}}$, could be used to constrain the HQE parameters, and the convergence of the HQE  is certainly questionable in the first case.  
Moreover, only two points are insufficient to determine the coefficients of  the HQE for the meson masses with useful precision: they could be pinned down in a much more effective way if one had the meson masses corresponding to heavy quarks with mass between the physical charm and $b$-quark masses \cite{Gambino:2012rd}, $m_c$  and $m_b$, or even above $m_b$.  
In this work we employ LQCD as a virtual laboratory to compute these meson masses with good accuracy. 

We have performed a precise lattice computation of PS and V meson masses for heavy-quark masses ranging from the physical charm mass up to $\simeq 4$ times the physical b-quark mass, using the gauge configurations generated by the European Twisted Mass Collaboration (ETMC) with $N_f = 2+1+1$ dynamical quarks at three values of the lattice spacing ($a \simeq 0.062, 0.082$ and $0.089$ fm) and with pion masses in the range $M_\pi \simeq 210 - 450$ MeV. 

Heavy-quark masses are simulated directly on the lattice up to $\simeq 3$ times the physical charm mass. 
The interpolation to the physical $b$-quark mass is obtained by adopting the {\it ETMC ratio} method \cite{Blossier:2009hg}, based on ratios of (spin-averaged) meson masses computed at nearby heavy-quark masses.
At variance with previous applications of the ETMC ratio method to $B$-physics \cite{Blossier:2009hg,Dimopoulos:2011gx,Carrasco:2013zta,Bussone:2016iua}, in this work we will adopt the heavy-quark mass defined in the kinetic scheme \cite{Bigi:1994em,Bigi:1996si} instead of the pole mass. 
The reason is that the kinetic mass is a short-distance mass free from the main renormalon ambiguities plaguing the pole mass \cite{Bigi:1994em,Bigi:1996si,Beneke:1994sw,Luke:1994xd,Martinelli:1995vj}.
This makes the choice of the kinetic scheme quite attractive for the analysis of inclusive $B$-meson decay data~\cite{Gambino:2011cq}.
The extrapolation to the physical pion mass and to the continuum limit yields $m_b^{\rm kin}(1~\mbox{GeV}) = 4.61 (20)$ GeV in agreement with the results of the OPE analysis of the inclusive semileptonic $B$-meson decays \cite{Gambino:2016jkc,Alberti:2014yda}.
Our result corresponds to $\overline{m}_b(\overline{m}_b) = 4.26 (18)$ GeV in the $\overline{MS}$ scheme, which is in agreement with the findings of Ref.~\cite{Bussone:2016iua} as well with other lattice determinations (see, e.g., Ref.~\cite{FLAG}).

Then, the ETMC ratio method is applied above the physical $b$-quark mass to provide heavy-light meson masses towards the static point. 
The lattice data are analyzed in terms of HQE, taking into account the anomalous dimension and the radiative corrections up to order ${\cal{O}}(\alpha_s^2)$ for the chromomagnetic operator~\cite{Heitger:2004gb,Guazzini:2007bu,Grozin:2007fh}.
The matrix elements of dimension-4 and dimension-5 operators, for which radiative corrections are known up to order ${\cal{O}}(\alpha_s^2)$, are extracted with a good precision, namely:
 \bea
        \label{eq:dim4}
        \overline{\Lambda} & = & 0.552~ (26) ~\mbox{GeV}  ~ , \\
        \label{eq:dim5_1}
        \mu_\pi^2 & = & 0.321 ~ (32)~\mbox{GeV}^2  ~ , \\
         \label{eq:dim5_2}
        \mu_G^2(m_b) & = & 0.253 ~ (25)~\mbox{GeV}^2 ~ .
  \eea
The data allows to estimate also the size of two combinations of the matrix elements of dimension-6 operators, for which radiative corrections are not yet available, namely:
 \bea
     \label{eq:dim6_1}
     \rho_D^3 - \rho_{\pi \pi}^3 - \rho_S^3 & = & 0.153 ~ (34) ~\mbox{GeV}^3 ~ , \\
     \label{eq:dim6_2}
     \rho_{\pi G}^3 + \rho_A^3 - \rho_{LS}^3 & = & -0.158 ~ (84) ~\mbox{GeV}^3 ~ .
 \eea
All the above HQE parameters, as well as the physical $c$- and $b$-quark masses, are highly correlated. 
Therefore the full covariance matrix is provided (see, later on, Tables \ref{tab:correlations_dim6}-\ref{tab:correlations_dim7}).
Our results (\ref{eq:dim4}-\ref{eq:dim6_2}), which are specific to the kinetic scheme, represent the first unquenched lattice determinations of the HQE parameters.

Ours is not the first attempt to extract the HQE parameters from the lattice.
In the past $\overline{\Lambda}$, $\mu_\pi^2$ and $\mu_G^2(m_b)$ have been estimated using quenched lattice QCD simulations \cite{Ewing:1995ih,Gimenez:1996nw,Gimenez:1996av,Kronfeld:2000gk}.
The lattice evaluations of Refs.~\cite{Gimenez:1996nw,Gimenez:1996av} were based on the subtraction of power divergencies generated by the mixing of the relevant operators with those of lower dimensionality. 
Instead, our approach is similar to the one adopted in Ref.~\cite{Kronfeld:2000gk} and, more recently, in Ref.~\cite{Komijani:2016jrh}.

The paper is organized as follows.
In section~\ref{sec:simulations} we describe the simulation details.
In section~\ref{sec:masses} we present the extraction of ground-state PS and V meson masses from the relevant two-point correlators.
In section~\ref{sec:ratio} we describe the basic features of the ETMC ratio method.
In section~\ref{sec:mbkin} we determine the $b$-quark mass in the kinetic scheme by analyzing the spin-averaged meson masses, while in section~\ref{sec:DeltaM} we analyze the hyperfine mass splitting and determine the mass difference ($M_{B^*} - M_B$).
In section \ref{sec:HQET} we apply the ETMC ratio method to calculate the PS and V meson masses beyond the physical $b$-quark mass and we perform their analysis in the HQE. 
Finally, section~\ref{sec:conclusions} contains our conclusions.

\section{Simulation details}
\label{sec:simulations}

The gauge ensembles used in this work have been generated by ETMC with $N_f = 2 + 1 + 1$ dynamical quarks, which include in the sea, besides two light mass-degenerate quarks, also the strange and the charm quarks \cite{Baron:2010bv,Baron:2011sf}.
The ensembles are the same adopted in Refs.~\cite{Carrasco:2014cwa,Bussone:2016iua} to determine the up, down, strange, charm and bottom quark masses. 

In the ETMC setup the Iwasaki action \cite{Iwasaki:1985we} for the gluons and the Wilson maximally twisted-mass action \cite{Frezzotti:2000nk,Frezzotti:2003xj,Frezzotti:2003ni} for the sea quarks are employed. 
Three values of the inverse bare lattice coupling $\beta$ and different lattice volumes are considered, as it is shown in Table \ref{tab:simudetails}, where the number of configurations analyzed ($N_{cfg}$) corresponds to a separation of $20$ trajectories.

At each lattice spacing different values of the light sea quark mass are considered, and the light valence and sea quark masses are always taken to be degenerate, i.e.~$m_\ell^{sea} = m_\ell^{val} = m_\ell$. 
In order to avoid the mixing of strange and charm quarks in the valence sector we adopt a non-unitary set up in which the valence strange and charm quarks are regularized as Osterwalder-Seiler fermions \cite{Osterwalder:1977pc}, while the valence up and down quarks have the same action of the sea.
Working at maximal twist such a setup guarantees an automatic ${\cal{O}}(a)$-improvement \cite{Frezzotti:2003ni,Frezzotti:2004wz}.
Quark masses are renormalized through the renormalization constant (RC) $Z_m = 1 / Z_P$, computed non-perturbatively using the RI$^\prime$-MOM scheme (see Ref.~\cite{Carrasco:2014cwa}).

We have simulated three values of the valence charm quark mass, which are needed to interpolate smoothly in the physical charm region.
The valence quark masses are in the following ranges: $3 m_{ud}^{phys} \lesssim m_\ell \lesssim 12 m_{ud}^{phys}$ and $0.7 m_c^{phys} \lesssim m_c \lesssim 1.1 m_c^{phys}$. 
In order to extrapolate up to the $b$-quark sector we have also considered seven values of the valence heavy-quark mass, $m_h$, in the range $1.1 m_c^{phys} \lesssim m_h \lesssim 3.3 m_c^{phys} \approx 0.8 m_b^{phys}$. 

The lattice scale is determined using the experimental value of $f_{\pi^+}$, while the physical up/down, strange and charm quark masses are obtained by using the experimental values for $M_\pi$, $M_K$ and $M_{D_s}$, respectively \cite{Carrasco:2014cwa}.

\begin{table}[hbt!]
{\footnotesize
\begin{center}
\begin{tabular}{||c|c|c|c||c|c|c||}
\hline
ensemble & $\beta$ & $V / a^4$ &$N_{cfg}$&$a\mu_\ell$& $a\mu_c$ & $a\mu_h > a \mu_c$ \\
\hline \hline
$A30.32$ & $1.90$ & $32^3\times 64$ & $150$ & $0.0030$ & $\{0.21256,$ & $\{0.34583, 0.40675,$ \\
$A40.32$ & & & $150$ & $0.0040$ & $~0.25000,$ & $~0.47840, 0.56267,$ \\
$A50.32$ & & & $150$ & $0.0050$ & $~~0.29404\}$ & $~0.66178, 0.77836,$ \\
\cline{1-1} \cline{3-5}
$A40.24$ & & $24^3\times 48 $ & $150$ & $0.0040$ & & $~0.91546\},$ \\
$A60.24$ & & & $150$ & $0.0060$ & & \\
$A80.24$ & & & $150$ & $0.0080$ & & \\
$A100.24$ & & & $150$ & $0.0100$ & & \\
\hline \hline
$B25.32$ & $1.95$ & $32^3\times 64$ & $150$ & $0.0025$ & $\{0.18705,$ & $\{0.30433, 0.35794,$ \\
$B35.32$ & & & $150$ & $0.0035$ & $~0.22000,$ & $~0.42099, 0.49515,$ \\
$B55.32$ & & & $150$ & $0.0055$ & $~~0.25875\}$ & $~0.58237, 0.68495,$ \\
$B75.32$ & & & $~75$ & $0.0075$ & & $~0.80561\}$ \\
\cline{1-1} \cline{3-5}
$B85.24$ & & $24^{3}\times 48 $ & $150$ & $0.0085$ & & \\
\hline \hline
$D15.48$ & $2.10$ & $48^3\times 96$ & $~90$ & $0.0015$ & $\{0.14454,$ & $\{0.23517, 0.27659,$ \\ 
$D20.48$ & & & $~90$& $0.0020$ & $~0.0150,$ & $~0.32531, 0.38262,$ \\
$D30.48$ & & & $~90$& $0.0030$ & $~~0.19995\}$ & $~0.45001, 0.52928,$ \\
                 & & &           &                 &                         & $~0.62252\}$\\
 \hline   
\end{tabular}
\end{center}
}
\caption{\it \small Values of the simulated valence-quark bare masses for the $15$ ETMC gauge ensembles with $N_f = 2+1+1$ dynamical quarks (see Ref.~\cite{Carrasco:2014cwa}). $N_{cfg}$ stands for the number of (uncorrelated) gauge configurations used in this work.}
\label{tab:simudetails}
\end{table}

In Ref.~\cite{Carrasco:2014cwa} eight branches of the analysis were considered. 
They differ in: 
\begin{itemize}
\item the continuum extrapolation adopting for the scale parameter either the Sommer parameter $r_0$ or the mass of a fictitious PS meson made up of strange(charm)-like quarks; 
\item the chiral extrapolation performed with fitting functions chosen to be either a polynomial expansion or a Chiral Perturbation Theory (ChPT) ansatz in the light-quark mass;
\item the choice between the methods M1 and M2, which differ by $O(a^2)$ effects, used to determine in the RI'-MOM scheme the mass RC $Z_m = 1 / Z_P$. 
\end{itemize}
In the present analysis we made use of the input parameters corresponding to each of the eight branches of Ref.~\cite{Carrasco:2014cwa}.
The central values and the errors of the input parameters, evaluated using bootstrap samplings with ${\cal{O}}(100)$ events, are collected in Table \ref{tab:8branches}.
Throughout this work all the results obtained within the above branches are averaged according to Eq.~(28) of Ref.~\cite{Carrasco:2014cwa}.

\begin{table}[htb!]
\begin{center} 
\begin{tabular}{||c|l|c|c|c|c|c||} 
\hline 
\multicolumn{1}{||c}{} & \multicolumn{1}{|c|}{$\beta$} & \multicolumn{1}{c|}{ $1^{st}$ } & \multicolumn{1}{c|}{ $2^{nd}$ } & \multicolumn{1}{c|}{ $3^{rd}$ } & \multicolumn{1}{c|}{ $4^{th}$ } \\ \hline    
               & 1.90 & 2.224(68) &2.192(75) &2.269(86)&2.209(84) \\ 
$a^{-1}({\rm GeV})$  & 1.95 & 2.416(63) &2.381(73) &2.464(85)&2.400(83) \\ 
               & 2.10 & 3.184(59) &3.137(64) &3.248(75)&3.163(75) \\ \cline{1-6} 
$m_{ud}^{phys}({\rm GeV})$ &&0.00372(13)&0.00386(17)&0.00365(10)&0.00375(13) \\ \cline{1-6} 
$m_c^{phys}$({\rm GeV})      && 1.183(34) &1.193(28)  &    1.177(25)&1.219(21) \\ \cline{1-6} 
\multicolumn{1}{||c}{}&\multicolumn{1}{|c|}{1.90} & \multicolumn{4}{c|}{ 0.5290(73) } \\ 
\multicolumn{1}{||c}{$Z_P$}&\multicolumn{1}{|c|}{1.95} & \multicolumn{4}{c|}{ 0.5089(34) } \\ 
\multicolumn{1}{||c}{}&\multicolumn{1}{|c|}{2.10} & \multicolumn{4}{c|}{ 0.5161(27) } \\ \hline   
\end{tabular}
\vspace{0.5cm}
\begin{tabular}{||c|l|c|c|c|c||} 
\hline 
\multicolumn{1}{||c}{}&\multicolumn{1}{|c|}{$\beta$} & \multicolumn{1}{c|}{ $5^{th}$ } & \multicolumn{1}{c|}{ $6^{th}$ } & \multicolumn{1}{c|}{ $7^{th}$ } & \multicolumn{1}{c||}{ $8^{th}$ }\\\hline    
               & 1.90 & 2.222(67)&2.195(75)   &2.279(89)&2.219(87) \\ 
$a^{-1}({\rm GeV})$  & 1.95 & 2.414(61)&2.384(73)   &2.475(88)&2.411(86) \\ 
                & 2.10 & 3.181(57)&3.142(64)   &3.262(79)&3.177(78) \\ \cline{1-6} 
$m_{ud}^{phys}({\rm GeV})$ &&0.00362(12)&0.00377(16)&0.00354(9)&0.00363(12) \\ \cline{1-6} 
$m_c^{phys}({\rm GeV})$      &&1.150(35) &1.158(27) &1.144(29) &1.182(19) \\ \cline{1-6}  
\multicolumn{1}{||c}{}&\multicolumn{1}{|c|}{1.90} &\multicolumn{4}{c||}{0.5730(42) } \\ 
\multicolumn{1}{||c}{$Z_P$}&\multicolumn{1}{|c|}{1.95} &\multicolumn{4}{c||}{ 0.5440(17) } \\ 
\multicolumn{1}{||c}{}&\multicolumn{1}{|c|}{2.10} &\multicolumn{4}{c||}{ 0.5420(10) } \\ \hline     
\end{tabular} 
\end{center} 
\vspace{-0.25cm}
\caption{\it The input parameters for the eight branches of the analysis of Ref.~\cite{Carrasco:2014cwa}. The renormalized quark masses and the RC $Z_P$ are given in the $\overline{\mathrm{MS}}$ scheme at a renormalization scale of 2 GeV. With respect to Ref.~\cite{Carrasco:2014cwa} the table includes an update of the values of the lattice spacing and, consequently, of all the other quantities.}
\label{tab:8branches}
\end{table}

\section{Extraction of ground-state meson masses}
\label{sec:masses}

The ground-state mass of pseudoscalar (PS) and vector (V) mesons can be determined by studying the appropriate two-point correlation functions at large (Euclidean) time distances $t$ from the source, viz.
 \be
    \label{eq:CPSt}
    C_{PS}(t) = \langle \sum_{\vec{x}} P_5(\vec{x}, t) P_5^{\dagger}(0,0) \rangle~ _{\overrightarrow{t \geq t_{\mathrm{min}}^{PS}}} ~
                        \frac{Z_{PS}}{2 M_{PS}} ~ \left[ e^{-M_{PS} t} + e^{-M_{PS} (T - t)} \right] ~ , 
 \ee
 \be
    \label{eq:CVt}
     C_V(t) = \frac{1}{3} \langle \sum_{i,\vec{x}} V_i(\vec{x}, t) V_i^{\dagger}(0,0) \rangle
                   ~ _{\overrightarrow{t \geq t_{\mathrm{min}}^V}} ~  \frac{Z_V}{2 M_V} ~ \left[ e^{-M_V t} + e^{-M_V (T - t)} \right] ~ , 
 \ee
where $M_{PS(V)}$ is the PS(V) ground-state mass and $t_{\mathrm{min}}^{PS(V)}$ stands for the minimum time distance at which the PS(V) ground state can be considered well isolated.
In Eqs.~(\ref{eq:CPSt}-\ref{eq:CVt}) $V_i(x) = \bar{q}_1(x) \gamma_i q_2(x)$ and $P_5(x) = \bar{q}_1(x) \gamma_5 q_2(x)$ represent, respectively, the interpolating fields for V and PS mesons, made of two valence quarks $q_1$ and $q_2$ with bare masses $\mu_1$ and $\mu_2$.
We set opposite values for the Wilson parameters of the two valence quarks ($r_1 = - r_2$), because this choice guarantees that the cutoff effects on the PS mass are ${\cal{O}}[a^2 (\mu_1 + \mu_2)]$~\cite{Frezzotti:2003ni}.
In what follows we will consider the quark $q_1$ to be either in the charm region or above, i.e.~$q_1 = c, h$, while the quark $q_2$ is always taken to be a light quark with bare mass $\mu_\ell$ (see Table \ref{tab:simudetails}).

The PS(V) ground-state mass, $M_{PS(V)}$, can be determined from the plateau of the effective mass $M_{PS(V)}^{eff}(t)$ at large time distances, viz.
 \be
     M_{PS(V)}^{eff}(t) \equiv \mbox{arcosh}\left[ \frac{C_{PS(V)}(t - 1) + C_{PS(V)}(t + 1)}{2 C_{PS(V)}(t)} \right] 
     ~ _{\overrightarrow{t \geq t_{\mathrm{min}}^{PS(V)}}} ~ M_{PS(V)} ~ .
     \label{eq:Meff}
 \ee

The statistical accuracy of the meson correlators (\ref{eq:CPSt}-\ref{eq:CVt}) can be significantly improved by the use of the ``one-end" trick stochastic method~\cite{Foster:1998vw, McNeile:2006bz}, which employs spatial stochastic sources at a single time slice chosen randomly.
Besides the use of local interpolating quark fields, in the case of charm or heavier quarks it is a common procedure to adopt also Gaussian-smeared interpolating quark fields~\cite{Gusken:1989qx} in order to suppress faster the contribution of the excited states, leading to an improved projection onto the ground state at relatively small time distances. 
For the values of the smearing parameters we set $k_G = 4$ and $N_G = 30$. 
In addition, we apply APE-smearing  to the gauge links~\cite{Albanese:1987ds} in the interpolating fields with parameters $\alpha_{APE} = 0.5$ and $N_{APE} = 20$.

We have implemented smeared fields both in the source and in the sink. 
We have therefore evaluated two-point correlation functions corresponding to the four possible combinations generated by using local/smeared fields at source/sink, namely $C_{PS(V)}^{LL}(t)$, $C_{PS(V)}^{LS}(t)$, $C_{PS(V)}^{SL}(t)$ and $C_{PS(V)}^{SS}(t)$, where $L$ and $S$ denote local and smeared operators, respectively. 

\begin{figure}[htb!]
\includegraphics[width=8.15cm]{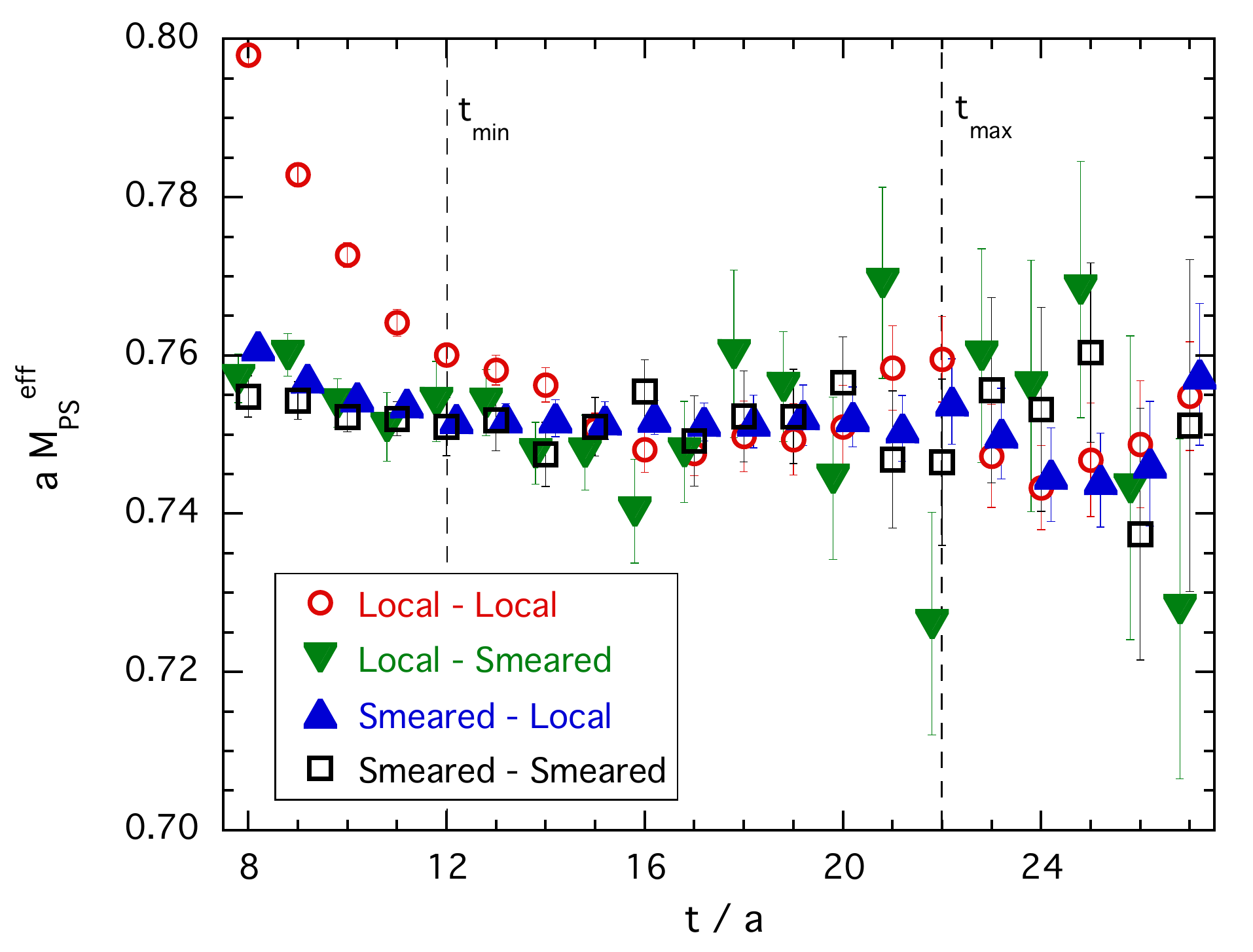}
\includegraphics[width=8.15cm]{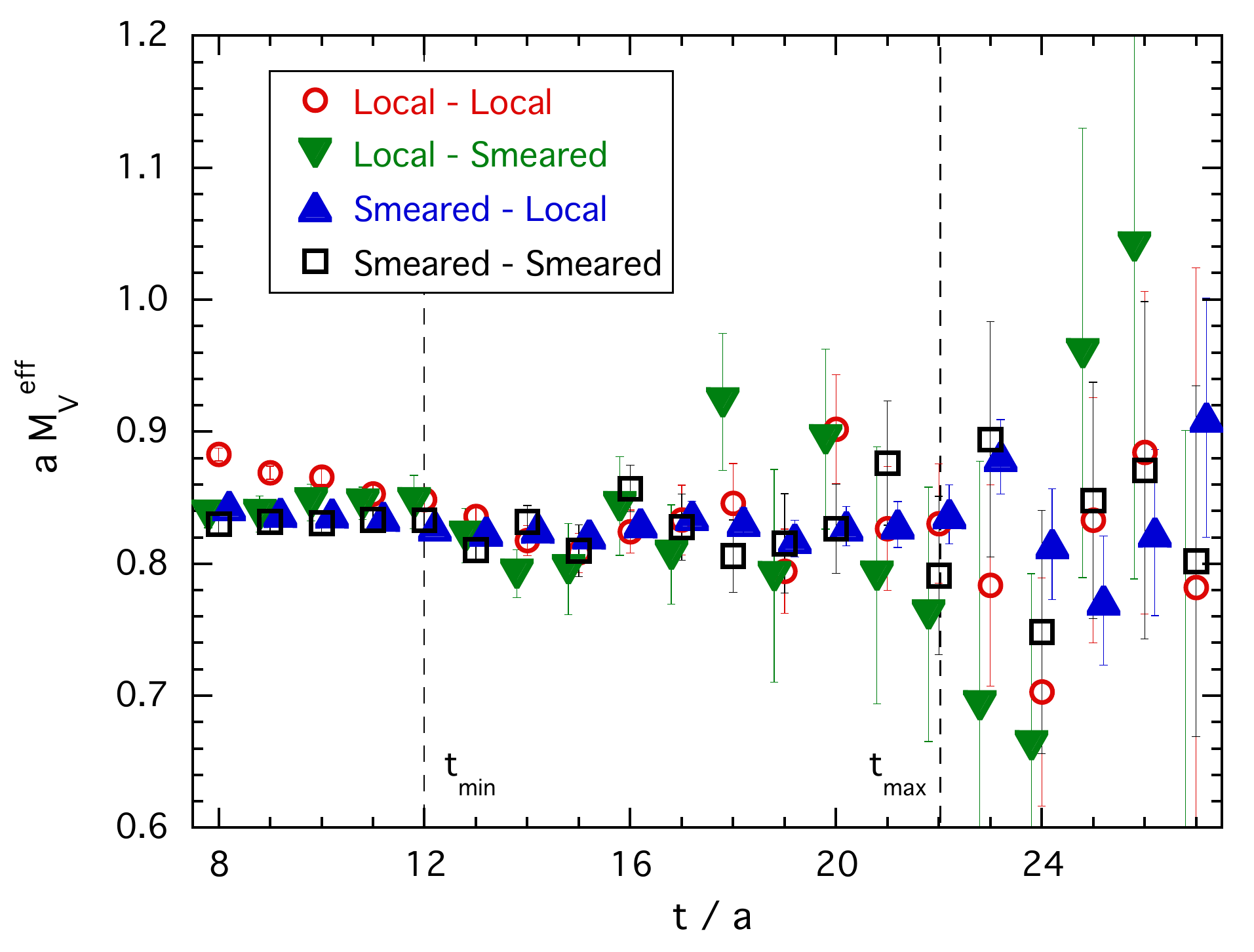}
\vspace{-1.0cm} 
\caption{\it \small Left panel: effective masses of the four correlators $C_{PS}^{LL}(t)$, $C_{PS}^{LS}(t)$, $C_{PS}^{SL}(t)$ and $C_{PS}^{SS}(t)$, calculated for a ($c \ell$) meson using Eq.~(\ref{eq:Meff}) in the case of the ETMC gauge ensemble B55.32 (corresponding to a pion mass equal to $\simeq 380$ MeV). Right panel: the same as in the left panel, but for the vector correlators $C_V^{LL}(t)$, $C_V^{LS}(t)$, $C_V^{SL}(t)$ and $C_V^{SS}(t)$.}
\label{fig:correlators}
\end{figure}

For the whole set of charm and heavier quark masses shown in Table \ref{tab:simudetails}, the $SL$ correlation functions exhibit the best signal to noise ratio, as it is illustrated in Fig.~\ref{fig:correlators} for a ($c \ell$) meson in the case of the gauge ensemble B55.32.

Thus, the $SL$ correlators have been used to extract the ground-state masses from the plateau of the effective mass (\ref{eq:Meff}) in the range $t_{\mathrm{min}}^{PS(V)} \leq t \leq t_{\mathrm{max}}^{PS(V)}$.
The stability of the extracted ground-state masses with respect to changes of both $t_{\mathrm{min}}^{PS(V)}$ and $t_{\mathrm{max}}^{PS(V)}$ has been studied and our choice of the values of $t_{\mathrm{min}}^{PS} = t_{\mathrm{min}}^V = t_{\mathrm{min}}$, $t_{\mathrm{max}}^{PS}$ and $t_{\mathrm{max}}^V$ in the charm sector is given in Table \ref{tab:tminmax}.

\begin{table}[htb!]
\begin{center}
\begin{tabular}{||c|c||c||c|c||}
\hline
$\beta$ & $V / a^4$ & $t_{\mathrm{min}} / a$ & $t_{\mathrm{max}}^{PS} / a$ & $t_{\mathrm{max}}^{V} / a$\\
\hline \hline
$1.90$ & $32^3 \times 64$ &$10$ &$30$ &$20$ \\
\cline{2-5}
            & $24^3 \times 48$ &$10$ &$20$ &$18$ \\
\hline \hline
$1.95$ & $32^3 \times 64$ &$12$ &$22$ &$20$ \\
\cline{2-5}
            & $24^3 \times 48$ & $12$ &$20$ &$18$ \\
\hline \hline
$2.10$ & $48^3 \times 96$ & $16$ &$44$ &$36$ \\ 
\hline   
\end{tabular}
\end{center}
\caption{\it Values of $t_{\mathrm{min}} = t_{\mathrm{min}}^{PS} = t_{\mathrm{min}}^V$, $t_{\mathrm{max}}^{PS}$ and $t_{\mathrm{max}}^V$ chosen to extract the ground-state signal from the effective mass (\ref{eq:Meff}), evaluated for heavy-light mesons with valence quark content ($c \ell$), using the SL correlators (i.e,~smeared quark fields in the source and local ones in the sink) in the case of the ETMC gauge ensembles of Table \ref{tab:simudetails}.}
\label{tab:tminmax}
\end{table}

\begin{figure}[htb!]
\includegraphics[width=8.15cm]{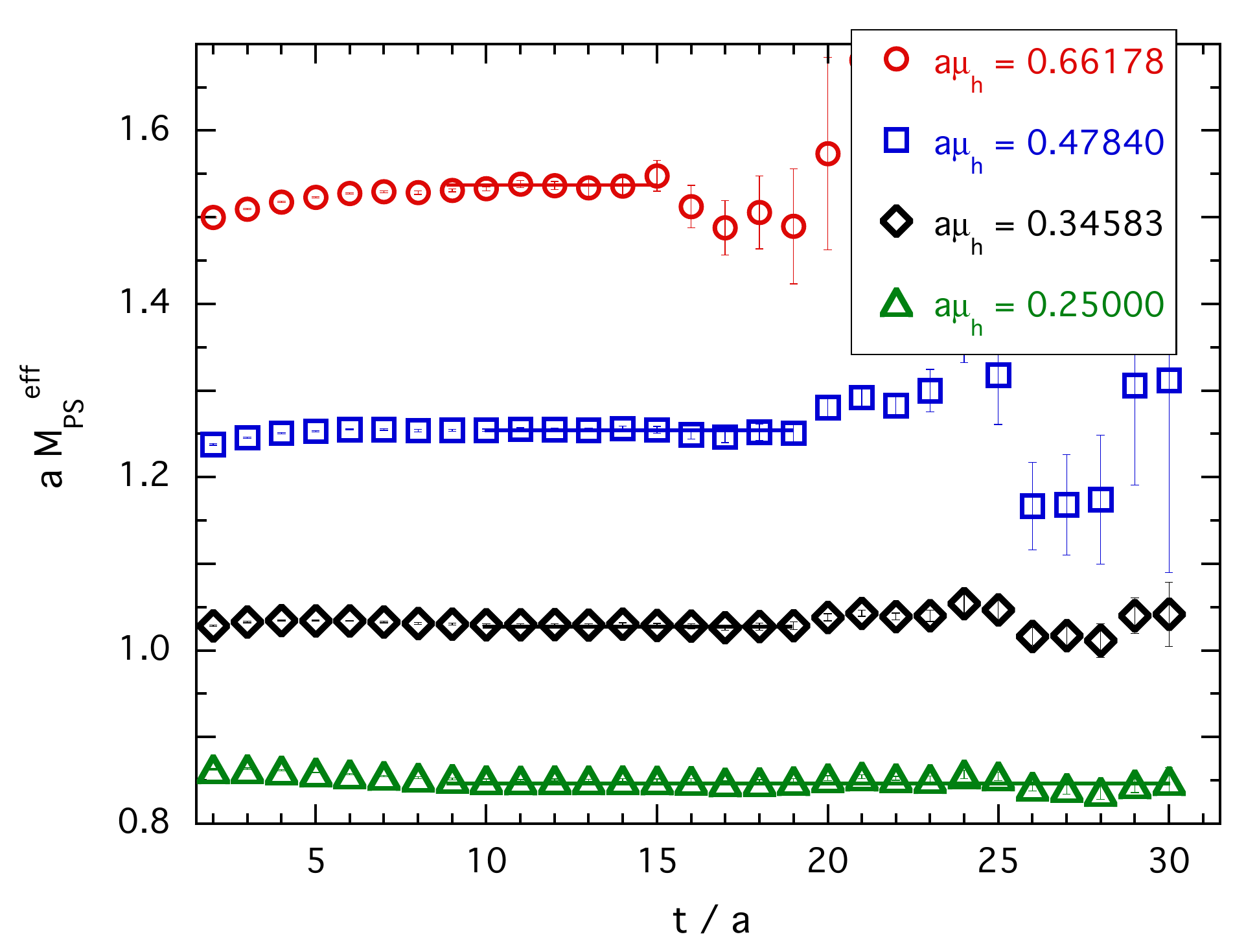}
\includegraphics[width=8.15cm]{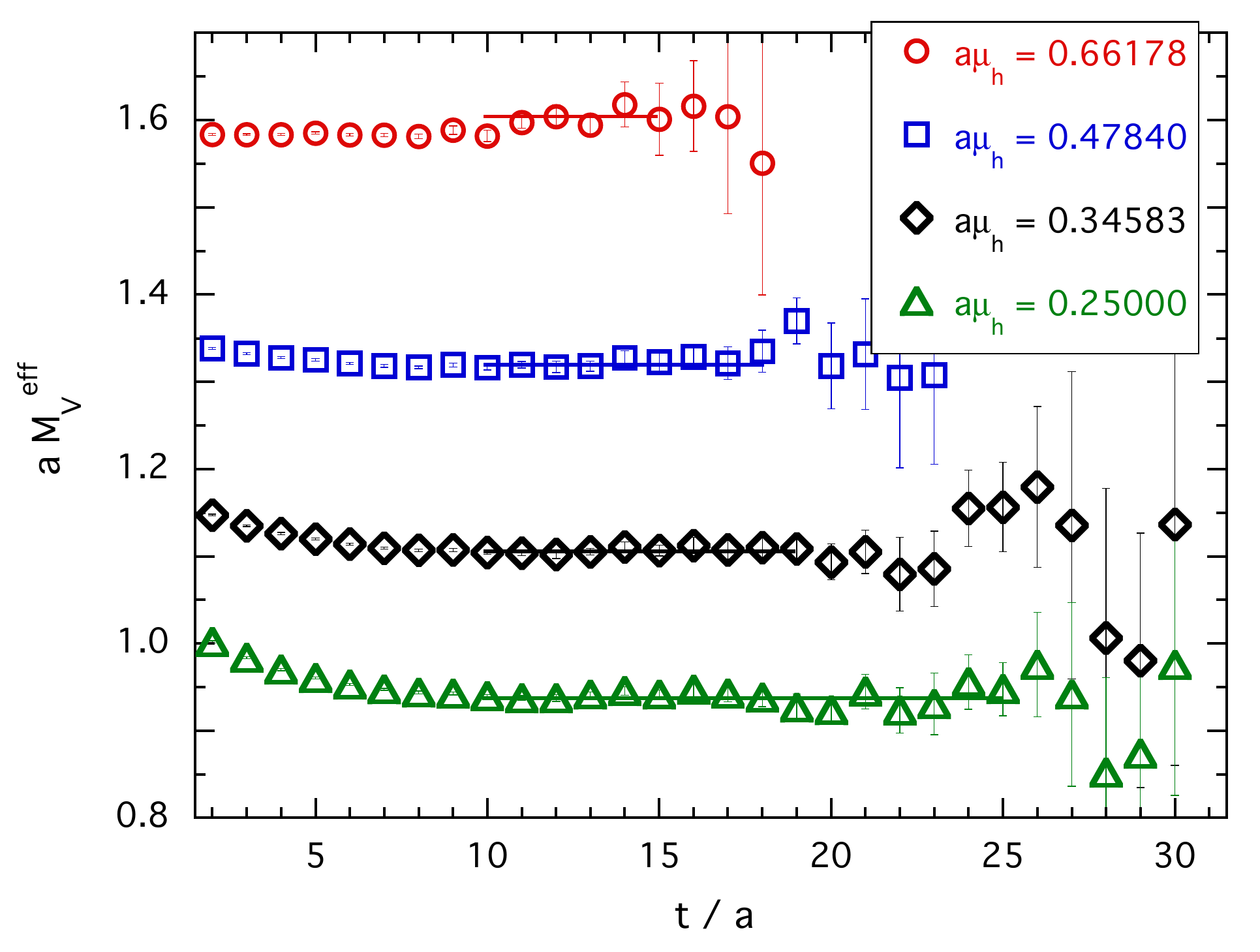}
\vspace{-1.0cm} 
\caption{\it \small Left panel: effective masses of the correlator $C_{PS}^{SL}(t)$ calculated for various ($h \ell$) meson using Eq.~(\ref{eq:Meff}) in the case of the ETMC gauge ensemble A40.32 (corresponding to a pion mass equal to $\simeq 320$ MeV). Right panel: the same as in the left panel, but for the vector correlator $C_V^{SL}(t)$. The solid lines identify the plateau region $t_{\mathrm{min}} \leq t \leq t_{\mathrm{max}}$ selected for each value of the heavy-quark mass.}
\label{fig:Meff-hl}
\end{figure}

The quality of the plateaux of the effective mass (\ref{eq:Meff}) is illustrated in Fig.~\ref{fig:Meff-hl} for a series of both PS and V heavy-light ($h \ell$) mesons in the case of the gauge ensemble A40.32.
It can be seen that the higher the heavy-quark mass the smaller the value adopted for $t_{\mathrm{max}}$, while the value chosen for $t_{\mathrm{min}}$ is independent on the heavy-quark mass.

We have checked our determination of the ground-state masses $M_{PS(V)}$ by employing an alternative method, namely the GEVP method of Ref.~\cite{Blossier:2009kd}, which is based on the simultaneous use of the four correlators $C_{PS(V)}^{LL}(t)$, $C_{PS(V)}^{LS}(t)$, $C_{PS(V)}^{SL}(t)$ and $C_{PS(V)}^{SS}(t)$.
It turns out that the GEVP method provides ground-state masses in nice agreement with those determined directly from the effective mass of the SL correlators with a slightly larger uncertainty.
Finally we have also checked that the impact of increasing by two units the values adopted for $t_{\mathrm{min}}$ in Table \ref{tab:tminmax} on the extracted PS and vector meson masses is negligible within present statistical uncertainties.

\section{The ETMC ratio method}
\label{sec:ratio}

Since the lattice spacing  of the ETMC gauge ensembles does not allow to simulate directly a $b$-quark on the lattice, the determination of quantities in the beauty sector requires alternative strategies.
In this respect a very suitable method is represented by the ETMC ratio method, already applied in the $N_f = 2$ framework~\cite{Blossier:2009hg,Dimopoulos:2011gx,Carrasco:2013zta} as well as in the $N_f = 2 + 1 + 1$ case~\cite{Bussone:2016iua} to determine the mass of the $b$-quark, the leptonic decay constants and the bag parameters of $B_{(s)}$ mesons. 

The ETMC ratio method consists in three main steps. 
The first one is the calculation of the observable of interest at heavy quark masses around the charm scale, for which relativistic simulations are reliable with well controlled discretisation errors. 
In the second step appropriate ratios of the observable are evaluated at increasing values of the heavy quark mass up to a scale of $\approx 3$ times the charm quark mass (i.e.~around 3 GeV). 
The crucial point is that the static limit of the ratios is exactly known from Heavy Quark Effective Theory (HQET) arguments. 
The final step of the computation is a smooth interpolation of the lattice data from the charm region to the infinite mass point, so that the value of the observable at the $b$-quark or $B$-meson mass can be determined. 

The great computational advantage of the ratio method is that $B$-physics computations can be carried out using the same relativistic action setup with which the lighter quark computations are performed. 
Moreover an extra simulation at the static point limit is not necessary, while the exact information about it is automatically incorporated in the construction of the ratios of the observable.
It should also be stressed that the use of ratios greatly helps in reducing the discretisation errors. 

As already explained in the Introduction, we are interested in studying the heavy-quark mass dependence of the following meson mass combinations:
 \bea
     \label{eq:Mav}
     M_{av}(\widetilde{m}_h) & \equiv & \frac{M_{PS}(\widetilde{m}_h) + 3 M_V(\widetilde{m}_h)}{4} ~ , \\[2mm]
      \label{eq:DM}
     \Delta M(\widetilde{m}_h) & \equiv & M_V(\widetilde{m}_h) - M_{PS}(\widetilde{m}_h) ~ ,
 \eea
where $\widetilde{m}_h = m_h^{kin}(\mu_{soft})$ is the renormalized heavy-quark mass in the kinetic scheme \cite{Bigi:1996si} at a soft cutoff $\mu_{soft}$, which is chosen to be equal to $\mu_{soft} = 1$ GeV.
For the sake of clarity, in what follows the renormalized quark mass in the $\overline{MS}$ scheme at a renormalization scale $\mu$ will be denoted by $\overline{m}_h(\mu)$.

At variance with previous applications of the ETMC ratio method, in this work we will adopt the heavy-quark mass $\widetilde{m}_h$ defined in the kinetic scheme instead of the pole mass $m_h^{pole}$.
The main reason is that the relation between the pole mass and the bare lattice masses $\mu_h$ suffers in perturbation theory from infrared renormalon ambiguities of order $O(\Lambda_{QCD})$ \cite{Bigi:1994em,Bigi:1996si,Beneke:1994sw,Luke:1994xd,Martinelli:1995vj}.
By the same token also the HQE parameter $\overline{\Lambda}$, measuring the difference between the heavy-hadron and heavy-quark masses, is affected by renormalon uncertainties and the same applies to other HQE parameters.
The kinetic mass $\widetilde{m}_h$ offers a solution to the above problem by subtracting from the pole mass its infrared sensitive part \cite{Bigi:1996si,Czarnecki:1997sz}, leading to a short-distance mass and to HQE parameters free from renormalon ambiguities.

The relation between the simulated bare heavy-quark mass $a \mu_h$ (see Table \ref{tab:simudetails}) and the kinetic mass $\widetilde{m}_h$ can be obtained in three steps. 
First, using the values of the lattice spacing and of the RC $Z_P$ from Table \ref{tab:8branches}, one gets 
\be
    \overline{m}_h(\mbox{2 GeV}) = \frac{1}{Z_P ~ a} ~ (a \mu_h)
    \label{eq:mh_2GeV}
 \ee
Then the perturbative scale can be evolved from $\mu = 2$ GeV to the value $\mu = \overline{m}_h$ using $\rm N^3LO$ perturbation theory \cite{Chetyrkin:1999pq} with four quark flavors ($n_\ell = 4$) and $\Lambda_{QCD}^{Nf = 4} = 297 (8)$ MeV \cite{PDG}, obtaining in this way $\overline{m}_h(\overline{m}_h)$.
Finally, the relation between the kinetic mass $\widetilde{m}_h$ and the $\overline{MS}$ mass $\overline{m}_h(\overline{m}_h)$ is known up to two loops \cite{Gambino:2011cq}, namely
 \bea
       \widetilde{m}_h & = & \overline{m}_h(\overline{m}_h) \left\{ 1 + \frac{4}{3} \frac{\alpha_s(\overline{m}_h)}{\pi}  
                                           \left[1 - \frac{4}{3} x - \frac{1}{2} x^2 \right] + \left( \frac{\alpha_s(\overline{m}_h)}{\pi} \right)^2 
                                           \right. \nonumber \\
                                 & \cdot & \left. \left[ \frac{\beta_0}{24} (8 \pi^2 + 71) + \frac{35}{24} + \frac{\pi^2}{9} \mbox{ln}(2) - 
                                                 \frac{7 \pi^2}{12} -\frac{\zeta_3}{6} \right. \right. \nonumber \\
                                 & + & \left. \left. \frac{4}{27} x \left( 24 \beta_0 \mbox{ln}(2x) - 64 \beta_0 + 6 \pi^2 - 39 \right) \right. \right.
                                           \nonumber \\
                                 & + & \left. \left. \frac{1}{18} x^2 \left( 24 \beta_0 \mbox{ln}(2x) - 52 \beta_0 + 6 \pi^2 - 23 \right) \right. \right. \nonumber \\
                                 & - & \left. \left. \frac{32}{27} x^3 -\frac{4}{9} x^4 \right] + {\cal{O}}(\alpha_s^3) \right\} ~ ,
       \label{eq:mkin}
 \eea
where $x \equiv \mu_{soft} / \overline{m}_h(\overline{m}_h)$, $\beta_0 = (33 - 2 n_\ell) / 12$ and $\zeta_3 \simeq 1.20206$.
We remind the reader that in the limit $\mu_{soft} \to 0$ the kinetic mass $\widetilde{m}_h$ coincides with the heavy-quark pole mass $m_h^{pole}$.
Between the charm and bottom scales the ratio $\widetilde{m}_h /  \overline{m}_h(\overline{m}_h)$ varies in the range $0.8 - 1.1$ and may be subject to important higher-order corrections. 
In Section \ref{sec:HQET} we will take into account the ensuing theoretical uncertainty.
Even within the present ${\cal{O}}(\alpha_s^2)$ accuracy the uncertainty in the determination of $\widetilde{m}_h$ can be decreased by optimizing the choice of the $\overline{MS}$ scale in Eq.~(\ref{eq:mh_2GeV}): we leave this for future improvements.

\section{Determination of the $b$-quark mass}
\label{sec:mbkin}

We start by applying the ratio method to the quantity $M_{av}(\widetilde{m}_h)$ (see Eq.~(\ref{eq:Mav})).
To this end we construct a sequence of heavy-quark masses $\widetilde{m}_h^{(n)}$ such that every two successive quark masses have a common fixed ratio $\lambda$, i.e.~for $n = 2, 3, ...$
 \be
      \widetilde{m}_h^{(n)} = \lambda \widetilde{m}_h^{(n - 1)} ~ .
      \label{eq:lambda}
 \ee

\begin{figure}[htb!]
\centering{\includegraphics[width=15.5cm]{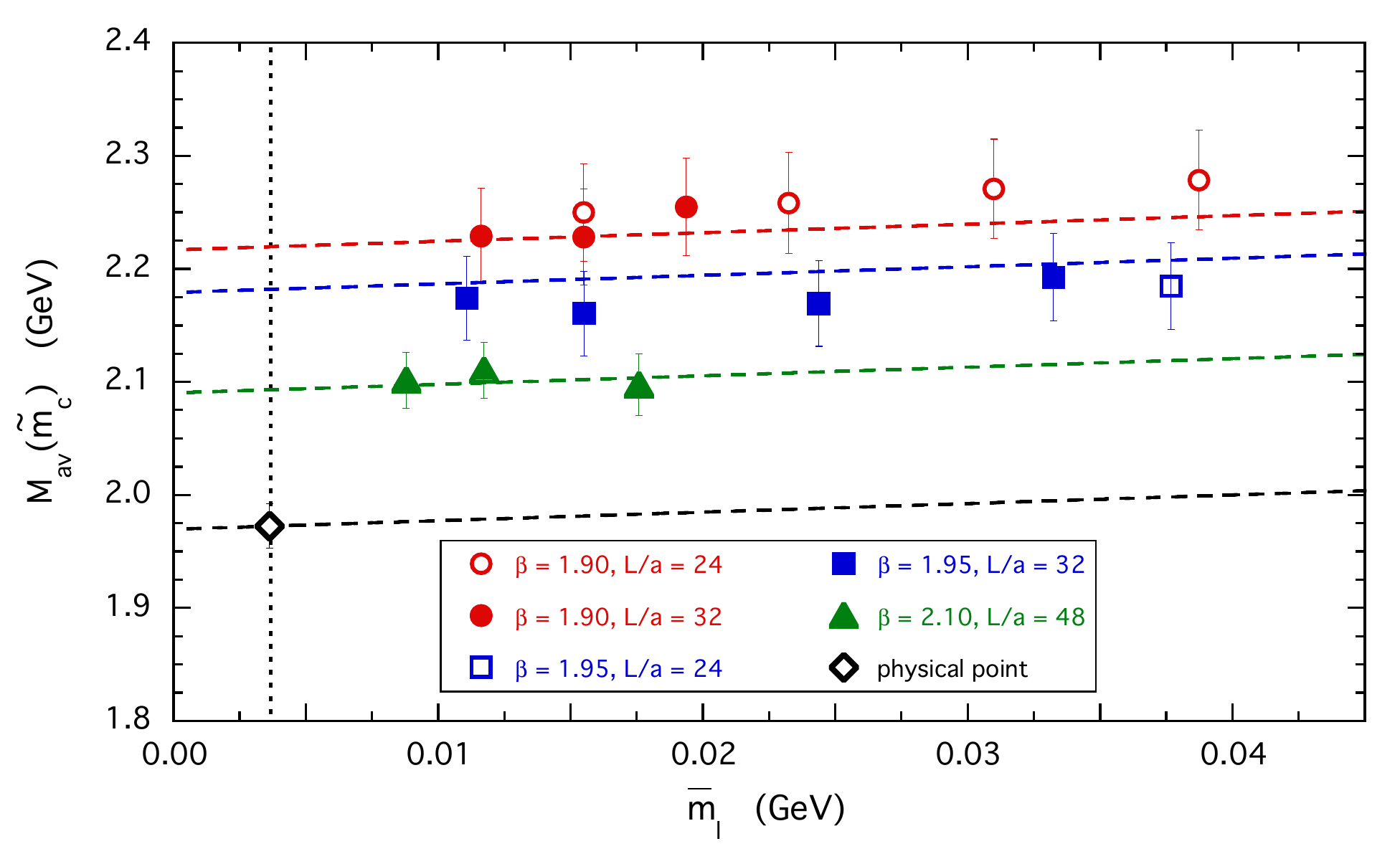}}
\vspace{-0.5cm}
\caption{\it \small The quantity $M_{av}(\widetilde{m}_h^{(1)}) = M_{av}(\widetilde{m}_c)$ versus the (renormalized) light-quark mass $\overline{m}_\ell = \overline{m}_\ell(\mbox{2 GeV})$ for the various ETMC gauge ensembles. The dashed lines are the results of a linear fit in both $\overline{m}_\ell$ and $a^2$ at each values of the lattice spacing and in the continuum limit. The diamond is the result at the physical light-quark mass $m_{ud}^{phys}$ (see Table \ref{tab:8branches}) in the continuum limit.}
\label{fig:Mav_charm}
\end{figure}

The series of masses starts at the physical charm quark mass $\widetilde{m}_h^{(1)} = \widetilde{m}_c = 1.219 (41)$ GeV corresponding to the result $\overline{m}_c(\mbox{2 GeV}) = 1.176 (36)$ GeV obtained in Ref.~\cite{Carrasco:2014cwa} using the experimental mass of the $D_s$-meson.
For each gauge ensemble the quantity $M_{av}(\widetilde{m}_c)$ can be safely computed by a smooth interpolation of the results corresponding to the subset of the bare quark masses in the charm region (see $a \mu_c$ in Table \ref{tab:simudetails}).
The lattice data for $M_{av}(\widetilde{m}_c)$ depend on the (renormalized) light-quark mass $\overline{m}_\ell$ and on the lattice spacing $a$.
They can be safely extrapolated to the physical pion mass (see $m_{ud}^{phys}$ in Table \ref{tab:8branches}) and to the continuum limit using a simple, combined linear fit in both $\overline{m}_\ell$ and $a^2$ (thanks to the automatic ${\cal{O}}(a)$-improvement of our lattice setup), as shown in Fig.~\ref{fig:Mav_charm}.
At the physical pion mass in the continuum limit we get $M_{av}^{phys}(\widetilde{m}_c) = 1.967 (25)$ GeV, which agrees with the experimental value $(M_D + 3 M_{D^*}) / 4 = 1.973$ GeV from PDG \cite{PDG} as well as with the result $M_{av}^{phys}(\widetilde{m}_c) = 1.975 (11)$ GeV based on the direct investigation of the $D^*$- to $D$-meson mass ratio of Ref.~\cite{Lubicz:2016bbi}.

Analogously, for each gauge ensemble the quantities $M_{av}(\widetilde{m}_h^{(n)})$ with $n = 2, 3, ...$ can be evaluated by interpolating the results corresponding to the subset of the bare heavy-quark masses (see $a \mu_h$ in Table \ref{tab:simudetails}).

\begin{figure}[htb!]
\includegraphics[width=8.15cm]{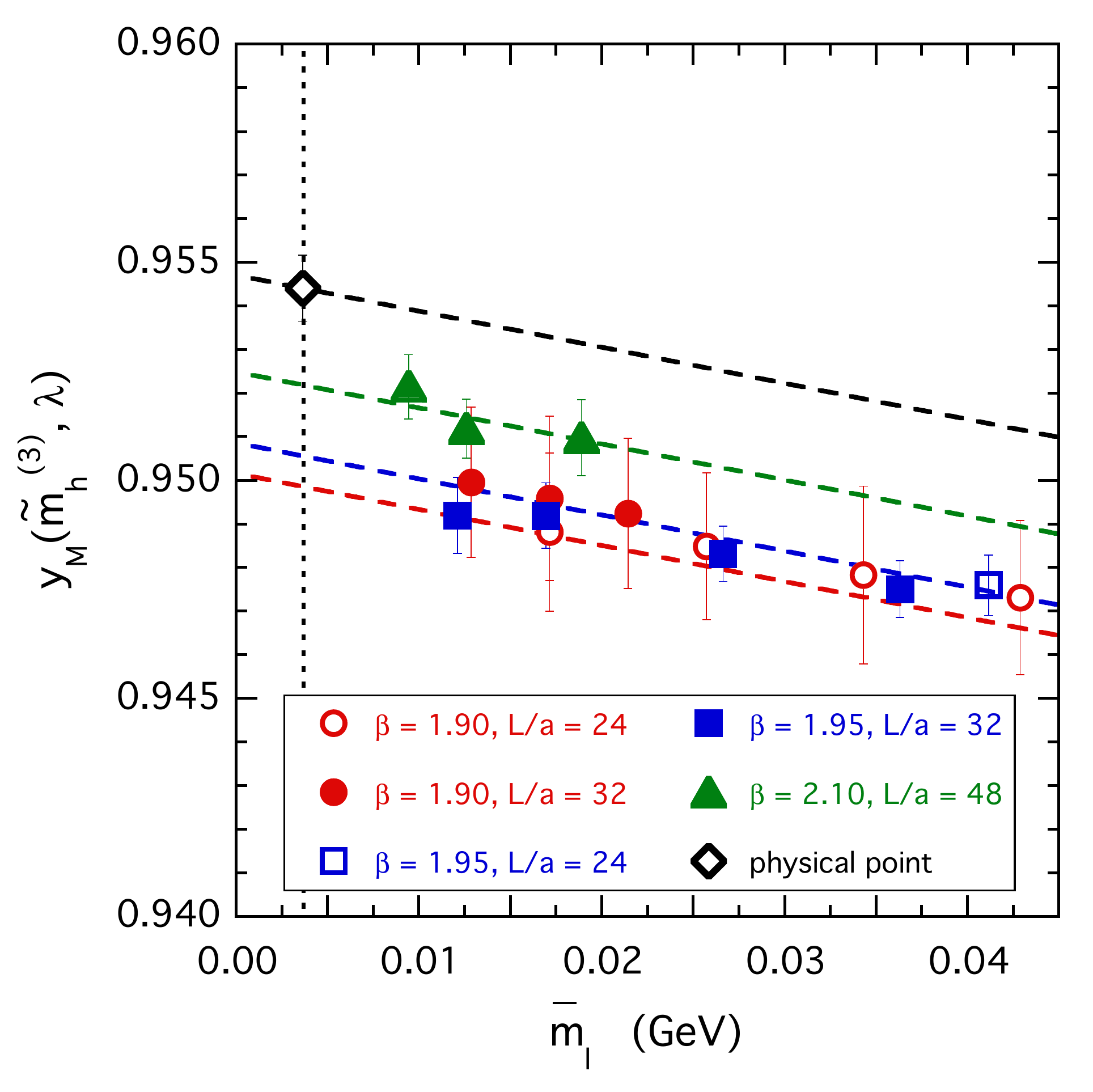}
\includegraphics[width=8.15cm]{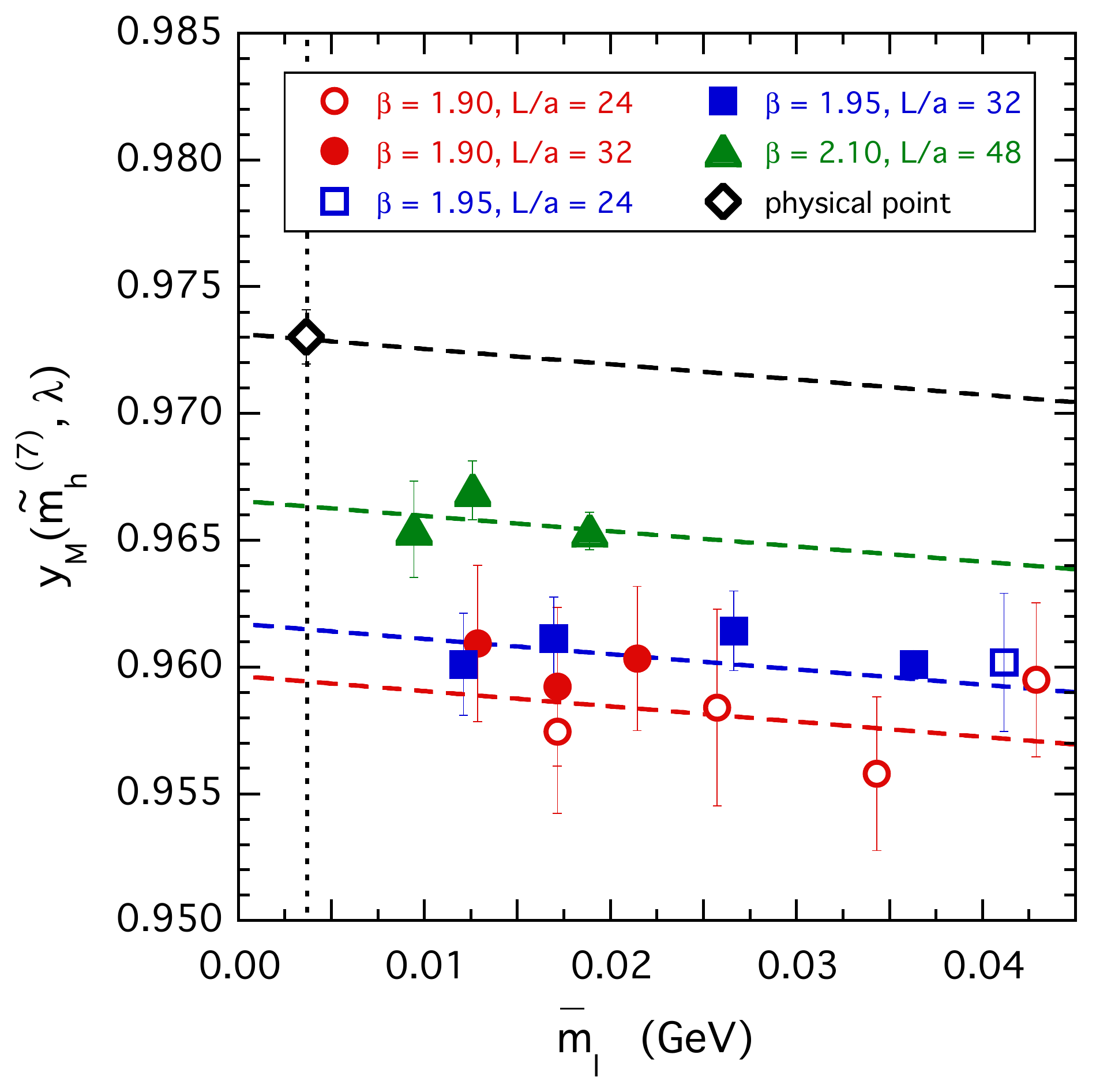}
\vspace{-1.0cm}
\caption{\it \small The ratios $y_M(\widetilde{m}_h^{(3)}, \lambda)$ (left panel) and $y_M(\widetilde{m}_h^{(7)}, \lambda)$ (right panel) versus the (renormalized) light-quark mass $\overline{m}_\ell = \overline{m}_\ell(\mbox{2 GeV})$ for the various ETMC gauge ensembles. The dashed lines are the results of a linear fit in both $\overline{m}_\ell$ and $a^2$. The diamonds correspond to the values $\overline{y}_M(\widetilde{m}_h^{(3)}, \lambda)$ and $\overline{y}_M(\widetilde{m}_h^{(7)}, \lambda)$, obtained at the physical light-quark mass $m_{ud}^{phys}$ (see Table \ref{tab:8branches}) in the continuum limit.}
\label{fig:yM}
\end{figure}

Then, we construct the following ratios
 \be
      y_M(\widetilde{m}_h^{(n)}, \lambda) = \frac{M_{av}(\widetilde{m}_h^{(n)})}{M_{av}(\widetilde{m}_h^{(n - 1)})} 
                                                                   \frac{\widetilde{m}_h^{(n - 1)}}{\widetilde{m}_h^{(n)}}
                                                                = \lambda^{-1} \frac{M_{av}(\widetilde{m}_h^{(n)})}{M_{av}(\widetilde{m}_h^{(n - 1)})}
      \label{eq:yM}
 \ee
with $n = 2, 3, ...$.
The advantage of considering the ratios (\ref{eq:yM}) is that discretization effects are suppressed even at the largest simulated value of the heavy-quark mass, as it is nicely illustrated in Fig.~\ref{fig:yM}. 

Each of the ratios $y_M(\widetilde{m}_h^{(n)}, \lambda)$ is therefore extrapolated to the physical pion mass and to the continuum limit using again a combined linear fit in both $\overline{m}_\ell$ and $a^2$, obtaining a value which will be denoted hereafter by $\overline{y}_M(\widetilde{m}_h^{(n)}, \lambda)$.
We have checked the possible impact of few systematics in the chiral and continuum limit extrapolations by considering either the inclusion of a quadratic term in the light-quark mass or the exclusion of the data at the coarsest lattice spacing ($\beta = 1.90$).
In both cases the differences of the extrapolated values $\overline{y}_M(\widetilde{m}_h^{(n)}, \lambda)$ are within the statistical uncertainties.

\begin{figure}[htb!]
\centering{\includegraphics[width=15.5cm]{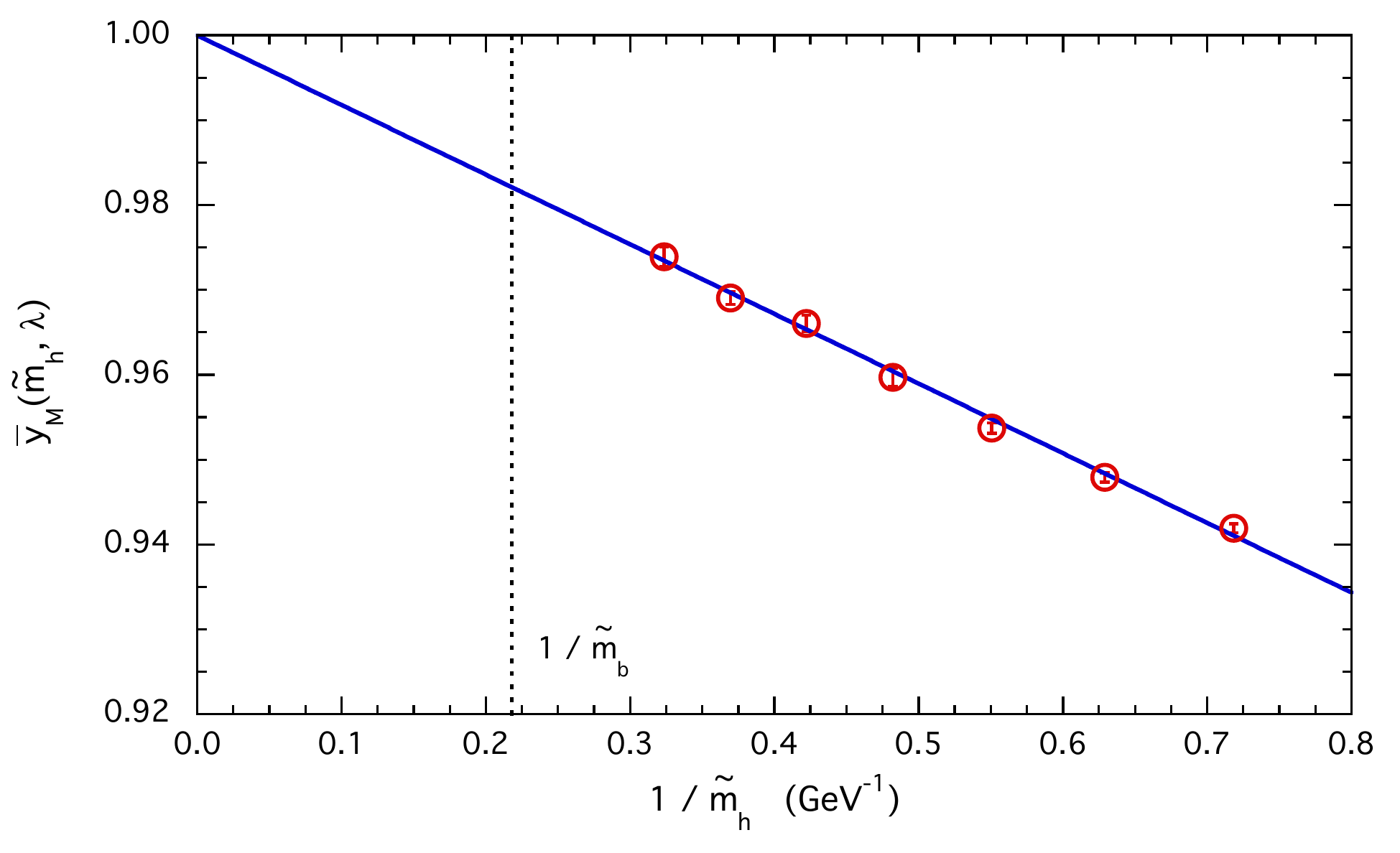}}
\vspace{-0.5cm}
\caption{\it \small Lattice data for the ratio $\overline{y}_M(\widetilde{m}_h, \lambda)$ versus the inverse heavy-quark mass $1 / \widetilde{m}_h$. The solid line is the result of the HQE-constrained fit (\ref{eq:yM_fit}) with $\epsilon_2 = 0$, taking into account the correlation matrix among the lattice points. The vertical dotted line corresponds to the position of the inverse physical $b$-quark mass $1 / \widetilde{m}_b$.}
\label{fig:yM_fit}
\end{figure}

In the static limit $\widetilde{m}_h \to \infty$ the HQE predicts
 \be
       \mbox{lim}_{\widetilde{m}_h \to \infty} ~ \frac{M_{av}(\widetilde{m}_h)}{\widetilde{m}_h} = 1 ~ ,
       \label{eq:yM_static}
 \ee
which implies $\mbox{lim}_{\widetilde{m}_h \to \infty} ~ \overline{y}_M(\widetilde{m}_h, \lambda) = 1$ for any value of $\lambda$.
Thus the $\widetilde{m}_h$-dependence of $\overline{y}_M$ can be described as a series expansion in terms of $1/\widetilde{m}_h$, namely
 \be
     \overline{y}_M(\widetilde{m}_h, \lambda) = 1 + \frac{\epsilon_1}{\widetilde{m}_h} +  \frac{\epsilon_2}{\widetilde{m}_h^2} + 
                                                                         {\cal{O}}\left( \frac{1}{\widetilde{m}_h^3} \right) ~ ,
     \label{eq:yM_fit}
 \ee
where the coefficients $\epsilon_{1,2}$ may depend upon $\lambda$.
The lattice data for the ratio $ \overline{y}_M(\widetilde{m}_h, \lambda)$ are shown in Fig.~\ref{fig:yM_fit} as a function of the inverse heavy-quark mass $1 / \widetilde{m}_h$.
It can be seen that a linear fit, i.e.~Eq.~(\ref{eq:yM_fit}) with $\epsilon_2 = 0$, is sufficient to fit the data taking into account the correlations between the lattice points. 
For each of the eight branches of the analysis (see Table \ref{tab:8branches}) the correlation matrix is constructed and the corresponding correlated $\chi^2$ variable is minimized.
The quality of the fit (\ref{eq:yM_fit}) with $\epsilon_2 = 0$ is illustrated in Fig.~\ref{fig:yM_fit}.

Finally, the chain equation
 \be
     \overline{y}_M(\widetilde{m}_h^{(2)}, \lambda) ~ \overline{y}_M(\widetilde{m}_h^{(3)}, \lambda) ~ ... ~ 
     \overline{y}_M(\widetilde{m}_h^{(K+1)}, \lambda) = \lambda^K \frac{M_{av}(\widetilde{m}_h^{(K+1)})}{M_{av}(\widetilde{m}_c)} ~ ,
     \label{eq:chain_M}
 \ee
in which the various factors in the l.h.s.~are evaluated through the fitting function (\ref{eq:yM_fit}), allows to determine the $b$-quark mass $\widetilde{m}_b$ by requiring that after $K$ (integer) steps the quantity $M_{av}(\widetilde{m}_h^{(K+1)})$ matches the experimental value $(M_B + 3 M_{B^*}) / 4 = 5.314$ GeV \cite{PDG}.
Then the $b$-quark mass $\widetilde{m}_b$ is directly given by $\widetilde{m}_b = \lambda^K ~ \widetilde{m}_c$.
In practice an iterative procedure should be applied in order to tune the value of the parameter $\lambda$ once the value of the integer $K$ is chosen.
Adopting $K = 10$ we find $\lambda = 1.1422 (10)$, which yields
 \be
     \widetilde{m}_b = 4.605 ~ (120)_{\rm stat} ~ (57)_{\rm syst}  ~ \mbox{GeV} = 4.605 ~ (132) ~ \mbox{GeV} ~ ,
     \label{eq:mb_kin}
\ee
where the systematic error comes from the eight branches of the input parameters of Table \ref{tab:8branches}. 
Translated in the $\overline{MS}$ scheme the result (\ref{eq:mb_kin}) corresponds to $\overline{m}_b(\overline{m}_b) = 4.257$ $(108)_{\rm stat}$ $(52)_{\rm syst}$ GeV $= 4.257$ $(120)$ GeV, which is well compatible with the ETMC determination $\overline{m}_b(\overline{m}_b) = 4.26$ $(10)$ GeV given in Ref.~\cite{Bussone:2016iua} and consistent with other lattice determinations within one standard deviation (see, e.g., the FLAG review~\cite{FLAG}).
The analysis of Ref.~\cite{Bussone:2016iua} shares the same ETMC gauge ensembles, but it differs in: ~ i) the use of the heavy-quark running mass $\overline{m}_h(\mbox{2 GeV})$ instead of the kinetic mass $\widetilde{m}_h$, ~ ii) a different definition of the ratios (\ref{eq:yM}), and ~ iii) the use of the experimental values of $B$- and $B_s$-meson masses instead of the spin-averaged $B$-meson mass to determine the $b$-quark mass.


Before closing the section we stress that the correlation $\rho$ between the determination (\ref{eq:mb_kin}) and the input value of the charm mass is $100 \%$, viz.
 \be
      \rho\left[\widetilde{m}_b, \widetilde{m}_c \right] = + 1 ~ .
      \label{eq:correlation_bc}
 \ee

\section{Analysis of the hyperfine meson mass splitting}
\label{sec:DeltaM}

In this Section we apply the ratio method to the hyperfine meson mass splitting $\Delta M(\widetilde{m}_h)$ (see Eq.~(\ref{eq:DM})).

\begin{figure}[htb!]
\centering{\includegraphics[width=15.5cm]{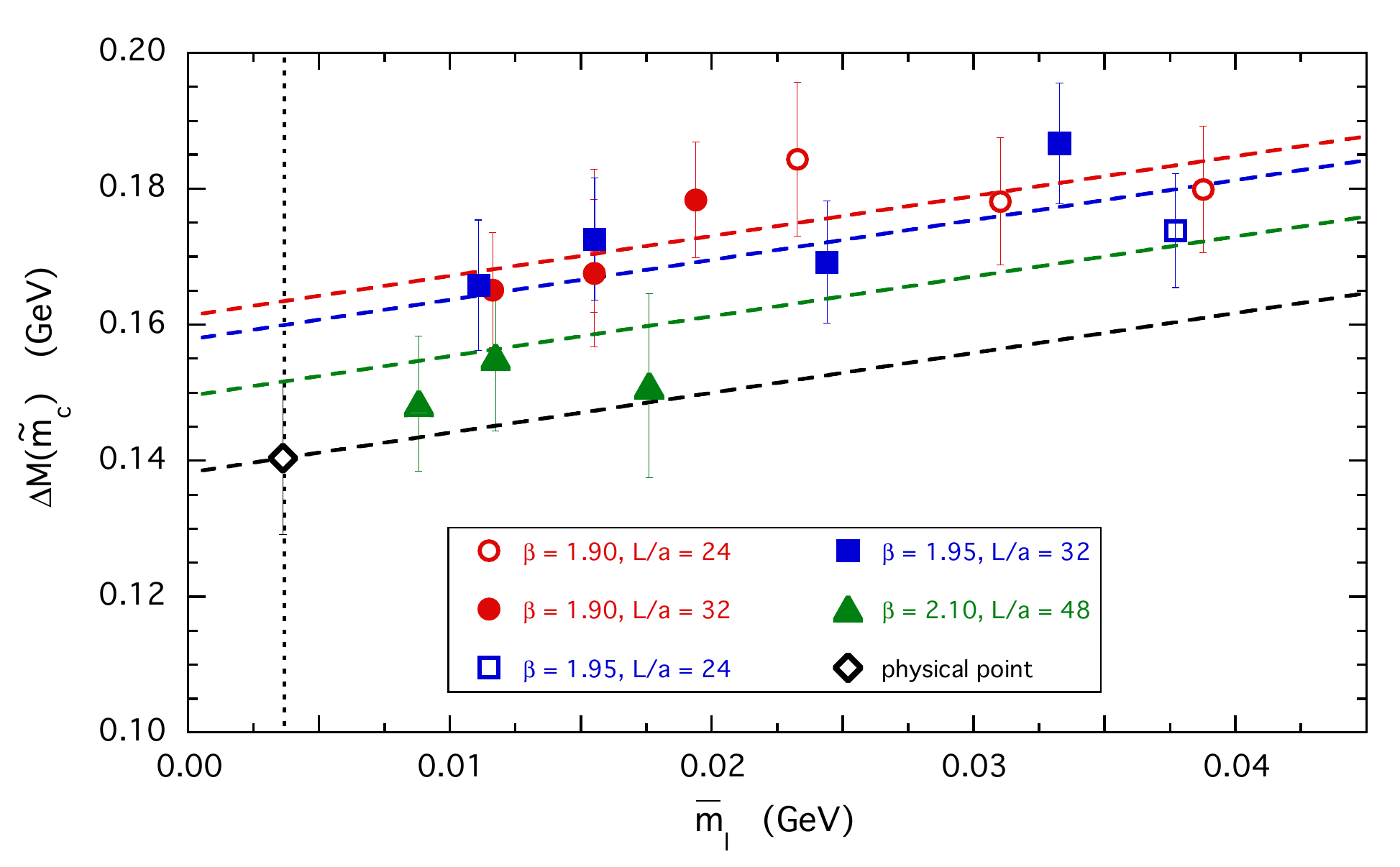}}
\vspace{-0.5cm}
\caption{\it \small The quantity $\Delta M(\widetilde{m}_h^{(1)}) = \Delta M(\widetilde{m}_c)$ versus the (renormalized) light-quark mass $\overline{m}_\ell = \overline{m}_\ell(\mbox{2 GeV})$ for the various ETMC gauge ensembles. The dashed lines are the results of a linear fit in both $\overline{m}_\ell$ and $a^2$ at each values of the lattice spacing and in the continuum limit. The black diamond is the result at the physical light-quark mass $m_{ud}^{phys}$ (see Table \ref{tab:8branches}) in the continuum limit.}
\label{fig:DM_charm}
\end{figure}

As in the case of the spin-averaged meson mass $M_{av}(\widetilde{m}_c)$, for each gauge ensemble the quantity $\Delta M(\widetilde{m}_c)$ at the triggering point $\widetilde{m}_c$ is computed by interpolating the results corresponding to the subset of the bare quark masses in the charm region (see $a \mu_c$ in Table \ref{tab:simudetails}).
Then the lattice data for $\Delta M_{av}(\widetilde{m}_c)$ are safely extrapolated to the physical pion mass and to the continuum limit using a  combined linear fit in both $\overline{m}_\ell$ and $a^2$, as illustrated in Fig.~\ref{fig:DM_charm}.

At the physical pion mass in the continuum limit we get $\Delta M^{phys}(\widetilde{m}_c) = 140 (11)$ MeV, which nicely agrees with the experimental value $M_{D^*} - M_D = 141.4$ MeV from PDG \cite{PDG} as well as with the result $M_{D^*} - M_D = 144 (15)$ MeV obtained in Ref.~\cite{Lubicz:2016bbi} from a direct investigation of the $D^*$- to $D$-meson mass ratio.

Analogously, for each gauge ensemble the quantities $\Delta M(\widetilde{m}_h^{(n)})$ with $n = 2, 3, ...$ are evaluated by interpolating the results corresponding to the subset of the bare heavy-quark masses (see $a \mu_h$ in Table \ref{tab:simudetails}).
We now consider the following ratios
 \bea
     y_{\Delta M}(\widetilde{m}_h^{(n)}, \lambda) & \equiv &  \frac{\widetilde{m}_h^{(n )}}{\widetilde{m}_h^{(n - 1)}}
         \frac{\Delta M(\widetilde{m}_h^{(n)})}{\Delta M(\widetilde{m}_h^{(n - 1)})} 
         \frac{c_G(\widetilde{m}_h^{(n - 1)}, \widetilde{m}_b)}{c_G(\widetilde{m}_h^{(n)}, \widetilde{m}_b)} \nonumber \\
         & = & \lambda \frac{\Delta M(\widetilde{m}_h^{(n)})}{\Delta M(\widetilde{m}_h^{(n - 1)})} 
         \frac{c_G(\widetilde{m}_h^{(n - 1)}, \widetilde{m}_b)}{c_G(\widetilde{m}_h^{(n)}, \widetilde{m}_b)} ~ ,
     \label{eq:yDM}
 \eea
where $c_G(\widetilde{m}_h, \widetilde{m}_b)$ is the short-distance Wilson coefficient that multiplies the matrix element of the HQET chromomagnetic operator renormalized in the $\overline{MS}$ scheme at the scale of the physical $b$-quark mass through a multiplicative RC, $Z_{CMO}(\overline{m}_b)$, viz.
 \be
     \mu_G^2(\overline{m}_b) \equiv Z_{CMO}(\overline{m}_b) \frac{\langle B | \bar{h}_v G_{\mu \nu} \sigma^{\mu \nu} h_v | B \rangle}
                                                        {2 \langle B | B \rangle}
     \label{eq:CMO}
 \ee
with $h_v$ being the field describing a heavy quark inside a hadron moving with velocity $v$.
Note that the ratio (\ref{eq:yDM}) is independent of the reference scale of the physical $b$-quark mass (see later Eq.~(\ref{eq:evolution})).

The coefficient $c_G$ is given by the product of three factors
 \be
     c_G = \overline{c}_G \cdot {\cal{R}} \cdot \frac{\widetilde{m}_h}{m_h^{pole}} ~ ,
      \label{eq:cGkin}     
 \ee
where $\overline{c}_G$ matches the HQE chromomagnetic operator with the corresponding one in QCD, ${\cal{R}}$ represents its running in the $\overline{MS}$ scheme and the factor $\widetilde{m}_h / m_h^{pole}$ is introduced to cancel the pole mass from the contribution of the chromomagnetic operator to the hyperfine splitting, improving in this way the convergence of the perturbative expansion.
An alternative method to achieve that has been presented in Refs.~\cite{Heitger:2004gb,Guazzini:2007bu}. 

\begin{figure}[htb!]
\centering{\includegraphics[width=15.5cm]{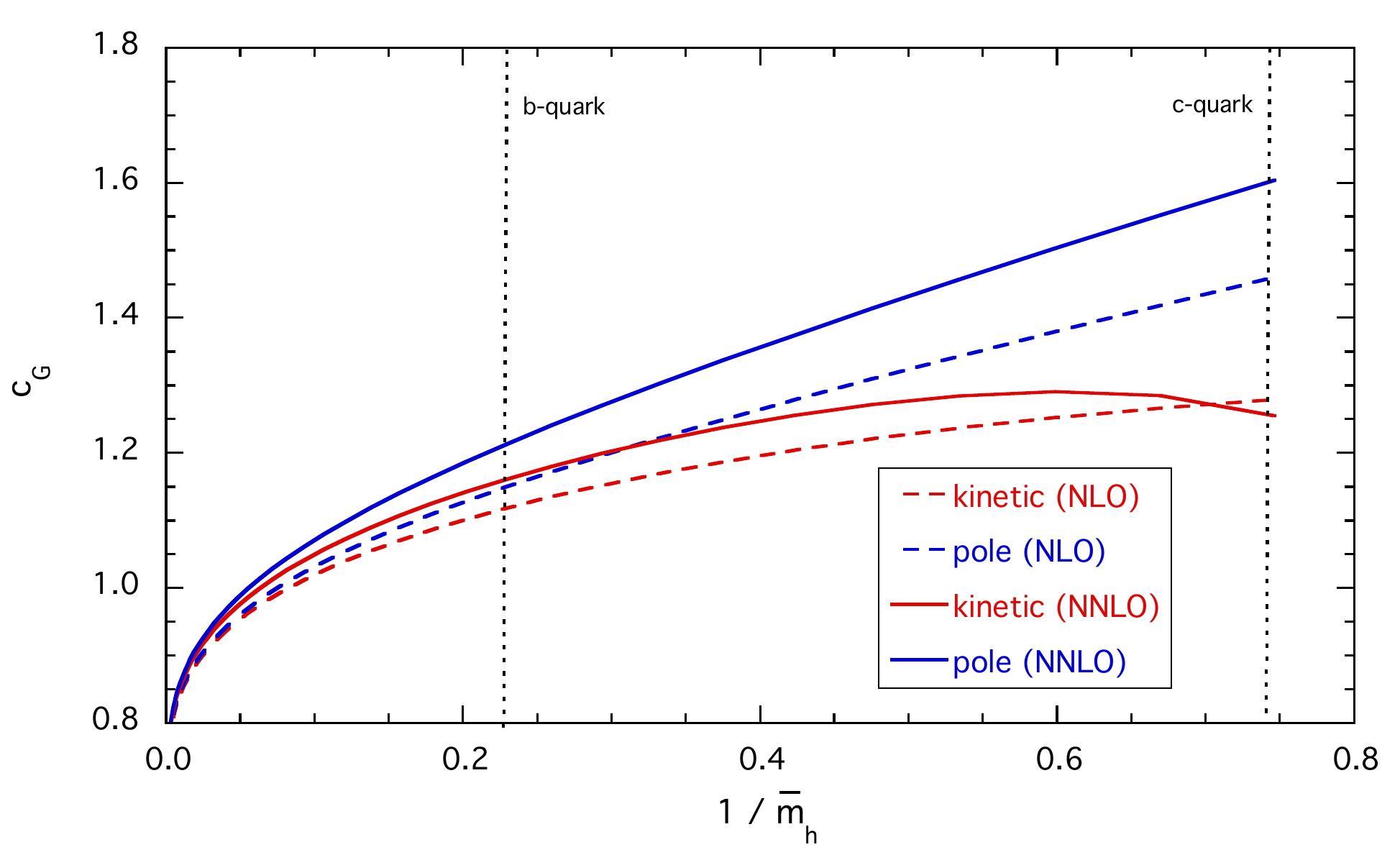}}
\vspace{-0.5cm}
\caption{\it \small The Wilson coefficient $c_G$ evaluated at orders ${\cal{O}}(\alpha_s)$ (dashed lines) and ${\cal{O}}(\alpha_s^2)$ (solid lines) in the kinetic scheme (red lines) and in the pole-mass scheme (blue lines), i.e.~using Eq.~(\ref{eq:cGkin_mkin}) with $\widetilde{x} \neq 0$ and $\widetilde{x} = 0$, respectively. The vertical dotted lines correspond to the locations of the inverse physical $b$-quark and $c$-quark masses.}
\label{fig:cG}
\end{figure}

The conversion coefficient $\overline{c}_G$ is known up to three loops in terms of $\alpha_s(m_h^{pole})$ \cite{Grozin:2007fh}. 
At two loops and in terms of $\alpha_s(\overline{m}_h)$ one gets 
 \be
     \overline{c}_G = 1 + \frac{13}{6} ~ \frac{\alpha_s(\overline{m}_h)}{\pi} + (11.4744 \beta_0 - 9.6584 ) ~ 
                                \left( \frac{\alpha_s(\overline{m}_h)}{\pi} \right)^2  + {\cal{O}}(\alpha_s^3) ~ .
      \label{eq:cGbar}
 \ee
The evolution factor ${\cal{R}}$ is given by
 \be
     {\cal{R}} = \left[ \frac{\alpha_s(\overline{m}_h)}{\alpha_s(\overline{m}_b)} \right]^{\frac{\gamma_0}{2 \beta_0}} ~ 
                      \frac{R(\overline{m}_h)}{R(\overline{m}_b)} ~ ,
     \label{eq:evolution}
 \ee
 with
 \be
     R(\overline{m}_h) \equiv 1 + r_1 \frac{\alpha_s(\overline{m}_h)}{\pi} + \frac{r_2 + r_1^2}{2}
                                             \left( \frac{\alpha_s(\overline{m}_h)}{\pi} \right)^2
     \label{eq:running}
 \ee
and
 \be
       r_1 = \frac{\gamma_0}{2 \beta_0} \left( \frac{\gamma_1}{\gamma_0} - \frac{\beta_1}{\beta_0} \right) ~ , \qquad
       r_2 = \frac{\gamma_0}{2 \beta_0} \left( \frac{\gamma_2}{\gamma_0} - \frac{\beta_1}{\beta_0} \frac{\gamma_1}{\gamma_0} - 
                 \frac{\beta_2}{\beta_0} + \frac{\beta_1^2}{\beta_0^2} \right) ~ . 
       \label{eq:r1&r2}
 \ee
In Eq.~(\ref{eq:r1&r2}) the parameters $\beta_i$ and $\gamma_i$ ($i = 0, 1, 2$) are respectively the loop coefficients of the QCD $\beta$ function and of the anomalous dimension $\gamma_{CMO}$ of the chromomagnetic operator, namely
 \bea
        \beta_0 & = & \left( 33 - 2 n_\ell \right) / 12 ~ , \\[2mm]
        \beta_1 & = & \left( 102 - \frac{38}{3} n_\ell \right) / 16 ~ , \\[2mm]
        \beta_2 & = & \left( 2857 - \frac{5033}{9} n_\ell + \frac{325}{27} n_\ell^2 \right) / 128
 \eea  
and \cite{Grozin:2007fh}
 \bea
        \gamma_0 & = & \frac{3}{2} ~ , \\[2mm]
        \gamma_1 & = & \left( 51 - \frac{13}{2} n_\ell \right) / 12 ~ , \\[2mm]
        \gamma_2 & = & 27 \left( \frac{\zeta_3}{8} + \frac{899}{1728} \right) + \frac{45}{48} \pi^2 - 
                                    \frac{n_\ell}{4} \left( 5\zeta_3 + \frac{57}{6} + \frac{5}{18} \pi^2 \right) - \frac{n_\ell^2}{48} ~ .
 \eea  
Moreover, from Eq.~(\ref{eq:mkin}) one has 
  \bea
       \frac{\widetilde{m}_h}{m_h^{pole}}  & = & 1 - \frac{4}{3} \frac{\alpha_s(\overline{m}_h)}{\pi} x \left( \frac{4}{3} + \frac{1}{2} x \right) 
                                                                          \nonumber \\
                                                               & + & \left( \frac{\alpha_s(\overline{m}_h)}{\pi} \right)^2 x \left[ \frac{4}{27} \left( 24 \beta_0 
                                                                         \mbox{ln}(2x) - 64 \beta_0 + 6 \pi^2 - 23 \right) \right. \nonumber \\
                                                               & + & \left.  \frac{1}{18} x \left( 24 \beta_0 \mbox{ln}(2x) - 52 \beta_0 + 6 \pi^2 - 7 \right) - 
                                                                         \frac{32}{27} x^2 - \frac{4}{9} x^3 \right] + {\cal{O}}(\alpha_s^3) ~ ,
       \label{eq:mkinpole}
  \eea
Introducing the variable $\tilde{x} \equiv \mu_{soft} / \widetilde{m}_h = x ~ \overline{m}_h / \widetilde{m}_h$ and taking into account that the values of the coupling constant $\alpha_s$ at the two scales $\overline{m}_h$ and $\widetilde{m}_h$ differ by terms of order ${\cal{O}}(\alpha_s^3)$ one finally obtains
 \bea
       c_G(\widetilde{m}_h, \widetilde{m}_b) & = & \frac{1}{R(\widetilde{m}_b)} \left[ \frac{\alpha_s(\widetilde{m}_h)}
                                                                             {\alpha_s(\widetilde{m}_b)} \right]^{\frac{\gamma_0}{2 \beta_0}}
                                                                             \left\{ 1 + \frac{\alpha_s(\widetilde{m}_h)}{\pi} \left[ \frac{13}{6} - 
                                                                             \frac{4}{3} \tilde{x} \left( \frac{4}{3} + \frac{1}{2} \tilde{x} \right) + r_1 \right] 
                                                                             \right. \nonumber \\
                                                                   & + & \left. \left( \frac{\alpha_s(\widetilde{m}_h)}{\pi} \right)^2 \left[ 11.4744 \beta_0 - 9.6584 +
                                                                             \frac{r_2 + r_1^2}{2} + \frac{13}{6} r_1 \right. \right. \nonumber \\
                                                                   & + & \left. \left. \frac{4}{27} \tilde{x} \left( 24 \beta_0 \mbox{ln}(2\tilde{x}) - 64 \beta_0 + 
                                                                             6 \pi^2 - 65 - 12 r_1 \right) \right. \right. \nonumber \\
                                                                   & + & \left. \left. \frac{1}{18}\tilde{x}^2 \left( 24 \beta_0 \mbox{ln}(2\tilde{x}) - 52 \beta_0 + 
                                                                            6 \pi^2 - \frac{73}{9} - 12 r_1 \right) \right. \right. \nonumber \\
                                                                   & - & \left. \left. \frac{32}{27} \widetilde{x}^3 - \frac{4}{9} \widetilde{x}^4 \right] + {\cal{O}}(\alpha_s^3) \right\} ~ .
      \label{eq:cGkin_mkin}
 \eea
The behavior of the coefficient $c_G$, calculated at orders ${\cal{O}}(\alpha_s)$ and ${\cal{O}}(\alpha_s^2)$, is shown in Fig.~\ref{fig:cG} in the case of the kinetic and pole-mass schemes, i.e.~Eq.~(\ref{eq:cGkin_mkin}) with $\widetilde{x} \neq 0$ and $\widetilde{x} = 0$, respectively.
It can be seen that the inclusion of the mass factor $\widetilde{m}_h / m_h^{pole}$ in Eq.~(\ref{eq:cGkin}) improves significantly the convergence of the perturbative expansion in agreement with expectations. 

\begin{figure}[htb!]
\includegraphics[width=8.15cm]{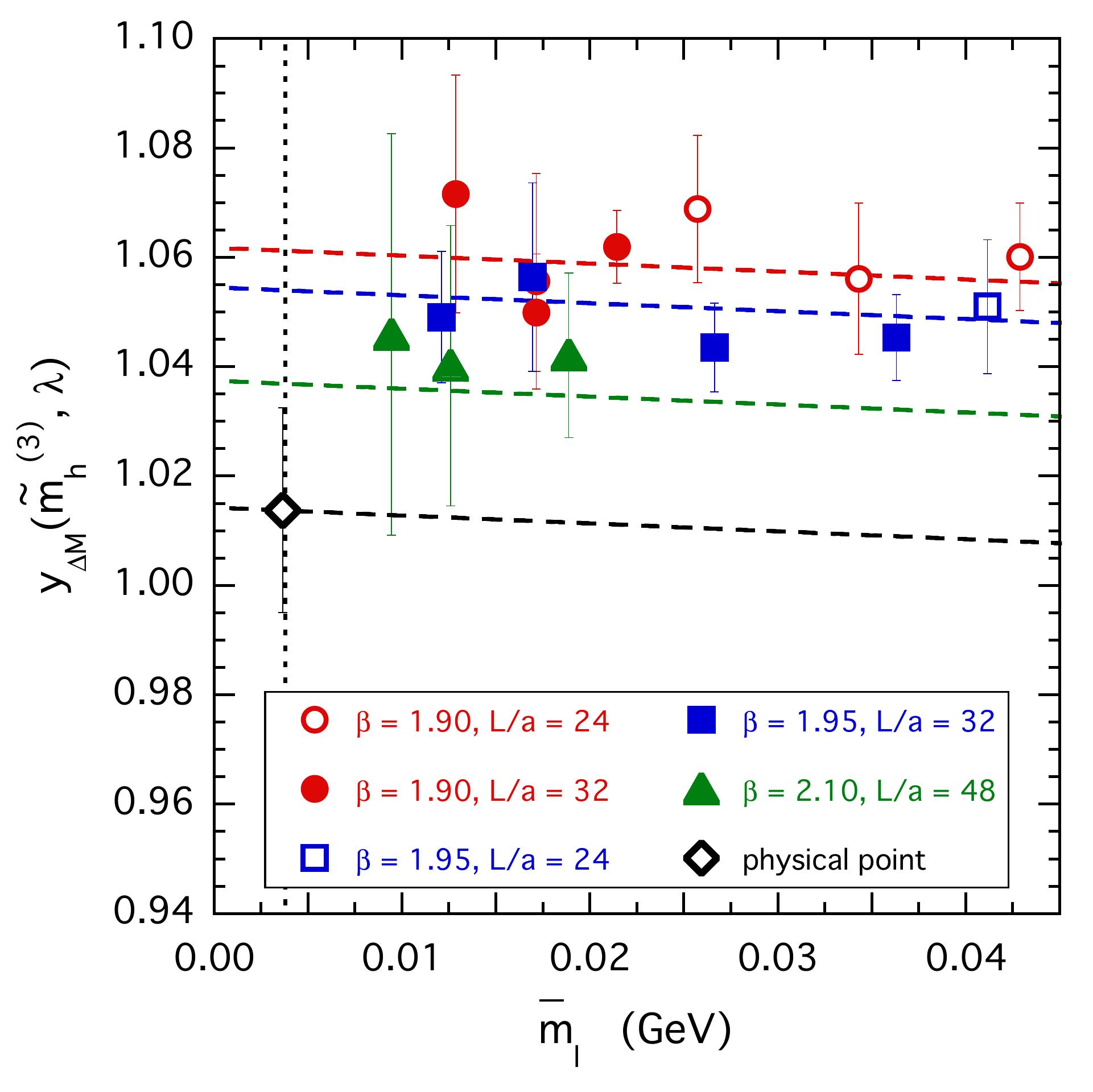}
\includegraphics[width=8.15cm]{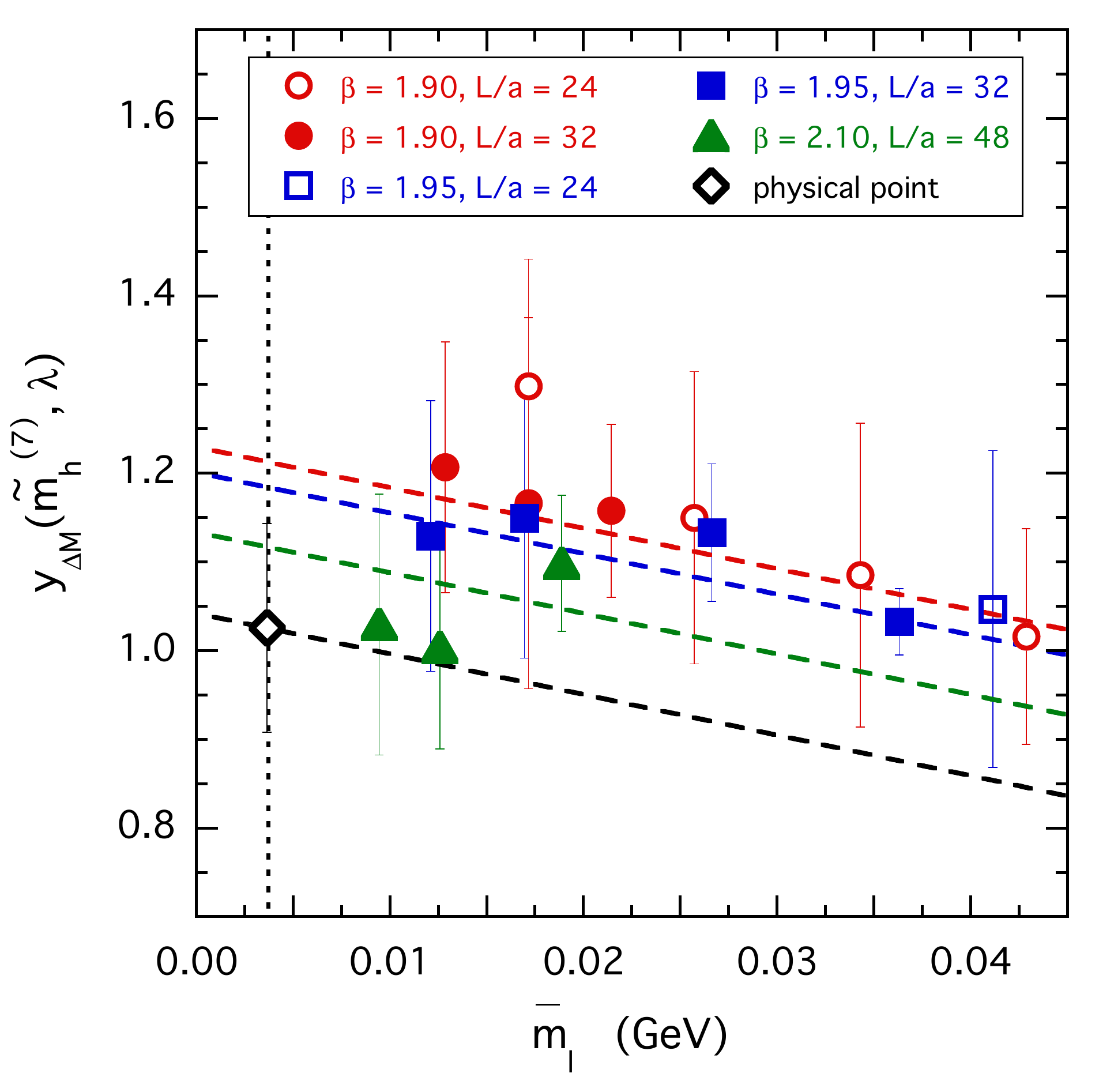}
\vspace{-1.0cm}
\caption{\it \small The ratios $y_{\Delta M}(\widetilde{m}_h^{(3)}, \lambda)$ (left panel) and $y_{\Delta M}(\widetilde{m}_h^{(7)}, \lambda)$ (right panel) versus the (renormalized) light-quark mass $\overline{m}_\ell = \overline{m}_\ell(\mbox{2 GeV})$ for the various ETMC gauge ensembles. The solid lines are the results of a linear fit in both $\overline{m}_\ell$ and $a^2$. The black dots correspond to the values $\overline{y}_{\Delta M}(\widetilde{m}_h^{(3)}, \lambda)$ and $\overline{y}_{\Delta M}(\widetilde{m}_h^{(7)}, \lambda)$, obtained at the physical light-quark mass $m_{ud}^{phys}$ (see Table \ref{tab:8branches}) in the continuum limit.}
\label{fig:yDM}
\end{figure}

The ratios (\ref{eq:yDM}) are extrapolated to the physical pion mass and to the continuum limit using a combined linear fit in both $\overline{m}_\ell$ and $a^2$, as shown in Fig.~\ref{fig:yDM}, obtaining a value which will be denoted hereafter by $\overline{y}_{\Delta M}(\widetilde{m}_h^{(n)}, \lambda)$.

\begin{figure}[htb!]
\centering{\includegraphics[width=15.5cm]{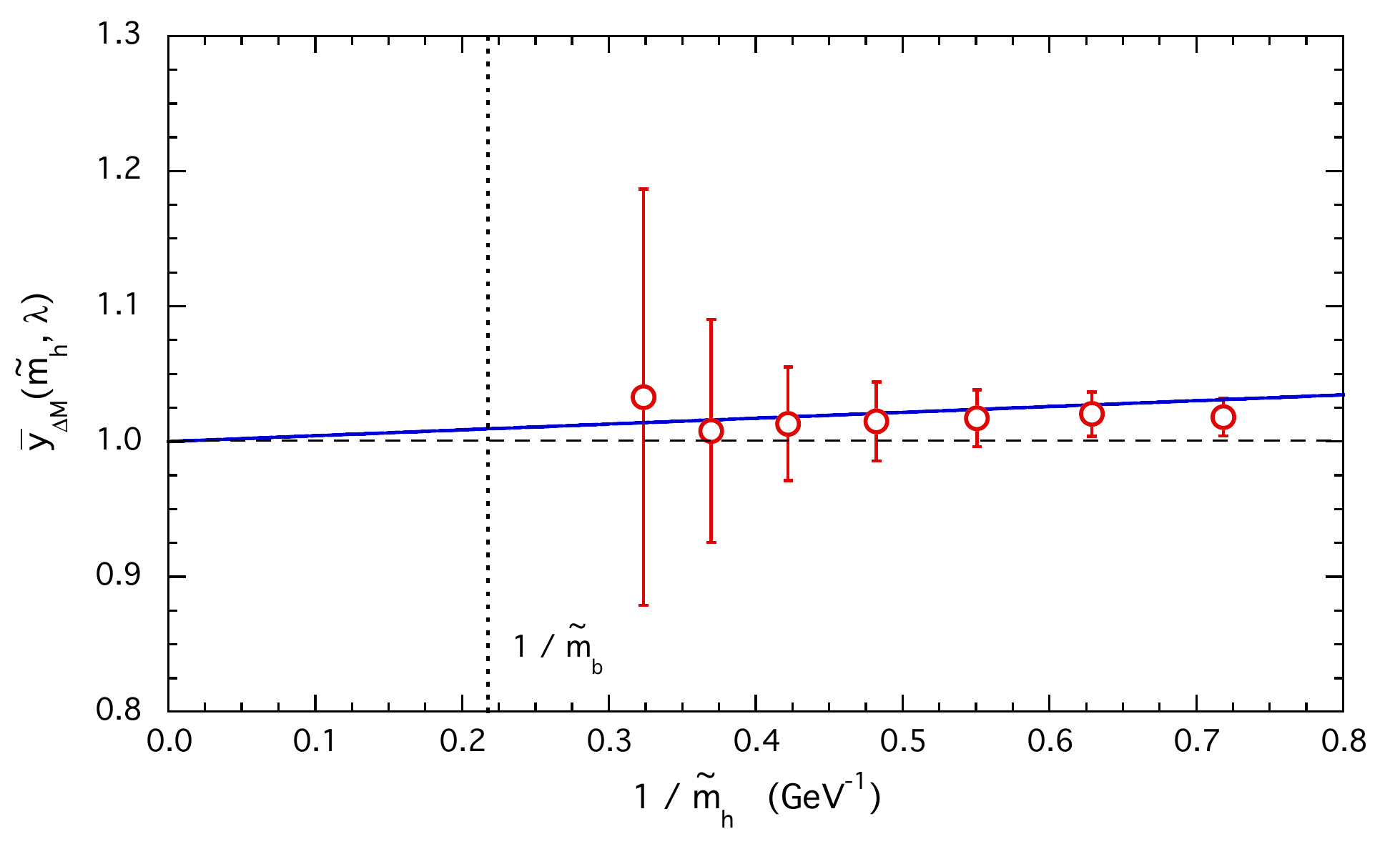}}
\vspace{-0.5cm}
\caption{\it \small Lattice data for the ratio $\overline{y}_{\Delta M}(\widetilde{m}_h, \lambda)$ versus the inverse heavy-quark mass $1 / \widetilde{m}_h$. The solid line is the result of the HQE-constrained fit (\ref{eq:yDM_fit}) with $\Delta \epsilon_2 = 0$, taking into account the correlation matrix among the lattice points. The vertical dotted line corresponds to the position of the inverse physical $b$-quark mass $1 / \widetilde{m}_b$.}
\label{fig:yDM_fit}
\end{figure}

In the static limit $\widetilde{m}_h \to \infty$ the HQE predicts
 \be
       \mbox{lim}_{\widetilde{m}_h \to \infty} ~ \widetilde{m}_h \frac{\Delta M(\widetilde{m}_h)}{c_G(\widetilde{m}_h, \widetilde{m}_b)} =
       \frac{2}{3} \mu_G^2(\widetilde{m}_b) ~ . 
       \label{eq:yDM_static}
 \ee
The HQE constraint (\ref{eq:yDM_static}) implies $\mbox{lim}_{\widetilde{m}_h \to \infty} ~ y_{\Delta M}(\widetilde{m}_h, \lambda) = 1$ for any value of $\lambda$.
Thus the $\widetilde{m}_h$-dependence of $\overline{y}_{\Delta M}$ can be described as a series expansion in terms of $1 / \widetilde{m}_h$, namely
 \be
     \overline{y}_{\Delta M}(\widetilde{m}_h, \lambda) = 1 + \frac{\Delta \epsilon_1}{\widetilde{m}_h} + 
                                                                                       \frac{\Delta \epsilon_2}{\widetilde{m}_h^2} + 
                                                                                       {\cal{O}}\left( \frac{1}{\widetilde{m}_h^3} \right) ~ ,
     \label{eq:yDM_fit}
 \ee
where the coefficients $\Delta \epsilon_{1,2}$ may depend upon $\lambda$.
The lattice data for the ratio $\overline{y}_{\Delta M}(\widetilde{m}_h, \lambda)$ are shown in Fig.~\ref{fig:yDM_fit} as a function of the inverse heavy-quark mass $1 / \widetilde{m}_h$.
As in the case of the spin-averaged ratios, a linear fitting function can be applied to the lattice data taking into account the correlations between the lattice points for each of the eight branches of the analysis.
The quality of the fit (\ref{eq:yDM_fit}) with $\Delta \epsilon_2 = 0$ is illustrated in Fig.~\ref{fig:yDM_fit}.

Using a chain equation analogous to Eq.~(\ref{eq:chain_M}) but expressed in terms of the ratios (\ref{eq:yDM}) and adopting the values of the parameters $\lambda$ and $K$ determined in the previous Section to reach the physical $b$-quark mass (\ref{eq:mb_kin}), we get for the hyperfine $B$-meson mass splitting the result $\Delta M(\widetilde{m}_b) = M_{B^*} - M_B = 40.2 (2.1)$ MeV, which is slightly below the experimental value $M_{B^*} - M_B = 45.42 (26)$ MeV \cite{PDG}, but improves the result $M_{B^*} - M_B = 41.2 (7.4)$ MeV of Ref.~\cite{Lubicz:2016bbi}, based on the direct investigation of the V to PS meson mass ratios.

Before closing the Section, we stress that throughout this work we have adopted four quark flavors ($n_\ell = 4$) and $\Lambda_{QCD}^{Nf = 4} = 297 (8)$ MeV \cite{PDG} also above the physical $b$-quark mass (\ref{eq:mb_kin}).
This is done mainly for consistency with the ETMC gauge ensembles used in this work and with the analyses of Ref.~\cite{Carrasco:2014cwa}, in which all the input parameters of Table \ref{tab:8branches} have been determined.

\section{Determination of the HQE expansion parameters}
\label{sec:HQET}

The chain equation (\ref{eq:chain_M}), as well as the analogous one in terms of the ratios (\ref{eq:yDM}), can be easily extended beyond the physical $b$-quark point using the fitting functions (\ref{eq:yM_fit}) with $\epsilon_2 = 0$ and (\ref{eq:yDM_fit}) with $\Delta \epsilon_2 = 0$.
In the case of the spin-averaged meson mass one obtains
 \bea
      \frac{M_{av}(\widetilde{m}_h^{(n)})}{\widetilde{m}_h^{(n)}} 
          & = & \frac{M_{av}(\widetilde{m}_c)}{\widetilde{m}_c} ~  \prod_{i = 2}^{n}
                    \overline{y}_M(\widetilde{m}_h^{(i)}, \lambda) ~ , \nonumber \\
          & = & \frac{M_{av}(\widetilde{m}_c)}{\widetilde{m}_c} ~  \prod_{i = 2}^{n}
                    \left[ 1 + \frac{\epsilon_1}{\lambda^{i-1} \widetilde{m}_c} \right] ~ ,
     \label{eq:Mavn}
 \eea
where $\widetilde{m}_h^{(n)} = \lambda^{n-1} ~ \widetilde{m}_c$, while for the hyperfine meson mass splitting one gets
 \bea
      \widetilde{m}_h^{(n)} \frac{\Delta M(\widetilde{m}_h^{(n)})}{c_G(\widetilde{m}_h^{(n)}, \widetilde{m}_b)} & = & 
          \widetilde{m}_c \frac{\Delta M(\widetilde{m}_c)}{c_G(\widetilde{m}_c, \widetilde{m}_b)} ~  \prod_{i = 2}^{n}
          \overline{y}_{\Delta M}(\widetilde{m}_h^{(i)}, \lambda) ~ , \nonumber \\
      & = & \widetilde{m}_c \frac{\Delta M(\widetilde{m}_c)}{c_G(\widetilde{m}_c, \widetilde{m}_b)} ~  \prod_{i = 2}^{n} 
          \left[ 1 + \frac{\Delta \epsilon_1}{\lambda^{i-1} \widetilde{m}_c} \right] ~ .
     \label{eq:DMn}
 \eea
 For values of $n > K+1$ Eqs.~(\ref{eq:Mavn}-\ref{eq:DMn}) provide V and PS heavy-meson masses beyond the physical $b$-quark point.
 In the static limit Eq.~(\ref{eq:Mavn}) implies
  \be
        Z_\infty \equiv \mbox{lim}_{\widetilde{m}_h \to \infty} \frac{M_{av}(\widetilde{m}_h)}{\widetilde{m}_h}  = 
        \frac{M_{av}(\widetilde{m}_c)}{\widetilde{m}_c} ~  \prod_{i = 2}^{\infty} \left[ 1 + \frac{\epsilon_1}{\lambda^{i-1} \widetilde{m}_c} 
        \right] ~ .
        \label{eq:Zstatic}
  \ee
The HQE predicts that $Z_\infty$ should be equal to unity.
Numerically we find $Z_\infty = 1.023 \pm 0.027$, which is well consistent with unity, but introduces a $\approx 3 \%$ uncertainty  in the static limit.
 In order to implement the exact condition $Z_\infty = 1$, for each bootstrap event we divide Eq.~(\ref{eq:Mavn}) by the definition (\ref{eq:Zstatic}) obtaining
  \be
       \frac{M_{av}(\widetilde{m}_h^{(n)})}{\widetilde{m}_h^{(n)}} = \frac{\prod_{i = 2}^{n} \left[ 1 + \frac{\epsilon_1}{\lambda^{i-1} 
           \widetilde{m}_c} \right]}{\prod_{i = 2}^{\infty} \left[ 1 + \frac{\epsilon_1}{\lambda^{i-1} \widetilde{m}_c} \right]} ~ .
       \label{eq:Mavn_final}
  \ee
 
\begin{figure}[htb!]
\includegraphics[width=15.5cm]{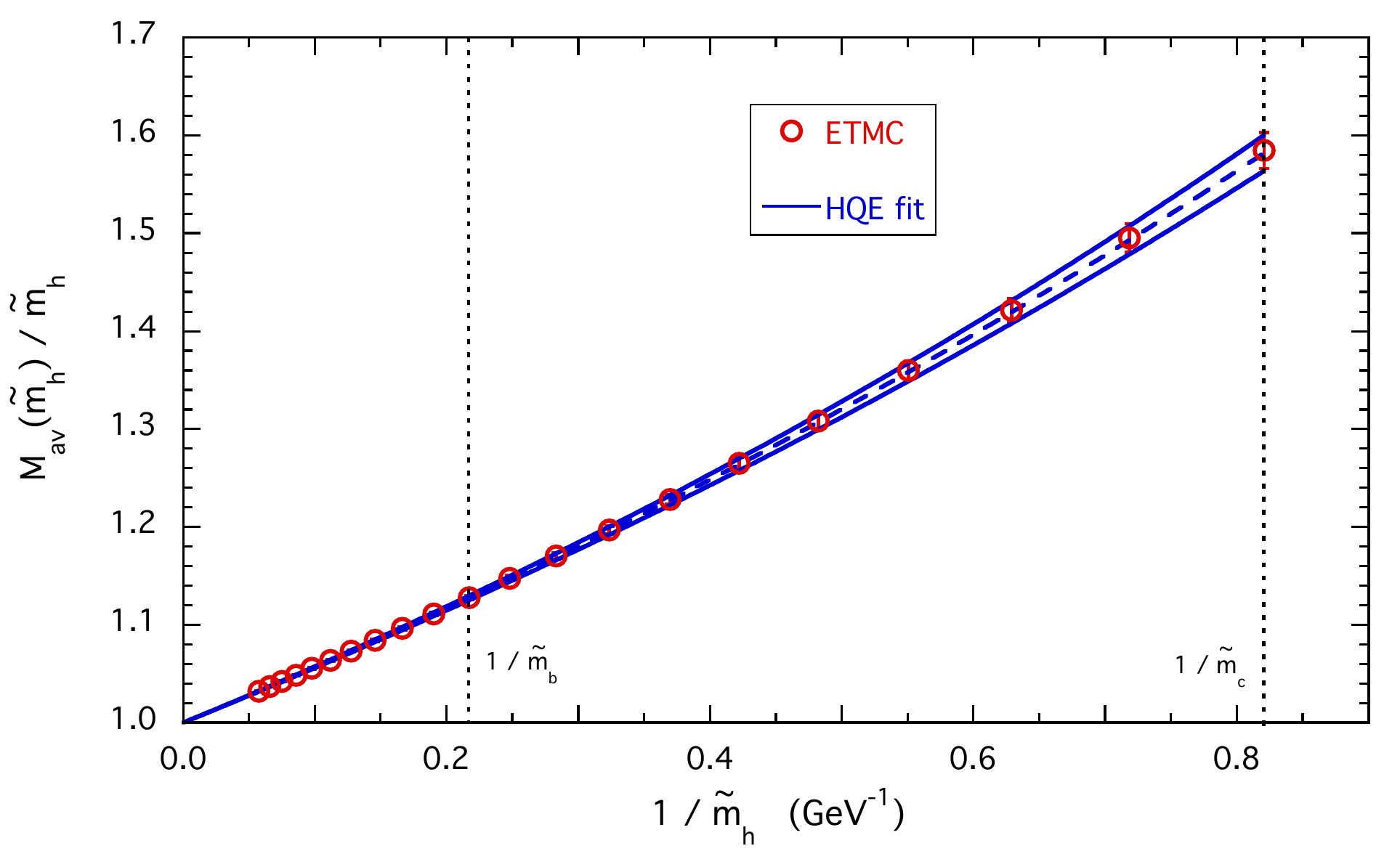}
\vspace{-0.5cm}
\caption{\it \small Lattice data for the quantity $M_{av}(\widetilde{m}_h) / \widetilde{m}_h$ (Eq.~(\ref{eq:Mavn_final})) versus the inverse heavy-quark mass $\widetilde{m}_h$. The dashed and solid lines are the results of the HQE fit (\ref{eq:M_HQET}) in which the correlation matrix between the lattice data is taken into account. The dashed line corresponds to the central values of the fits, while the solid lines represent one standard deviation. The vertical dotted lines correspond to the positions of the inverse physical $b$-quark and $c$-quark masses, $1 / \widetilde{m}_b$ and $1 / \widetilde{m}_c$.}
\label{fig:Mav}
\end{figure}

\begin{figure}[htb!]
\includegraphics[width=15.5cm]{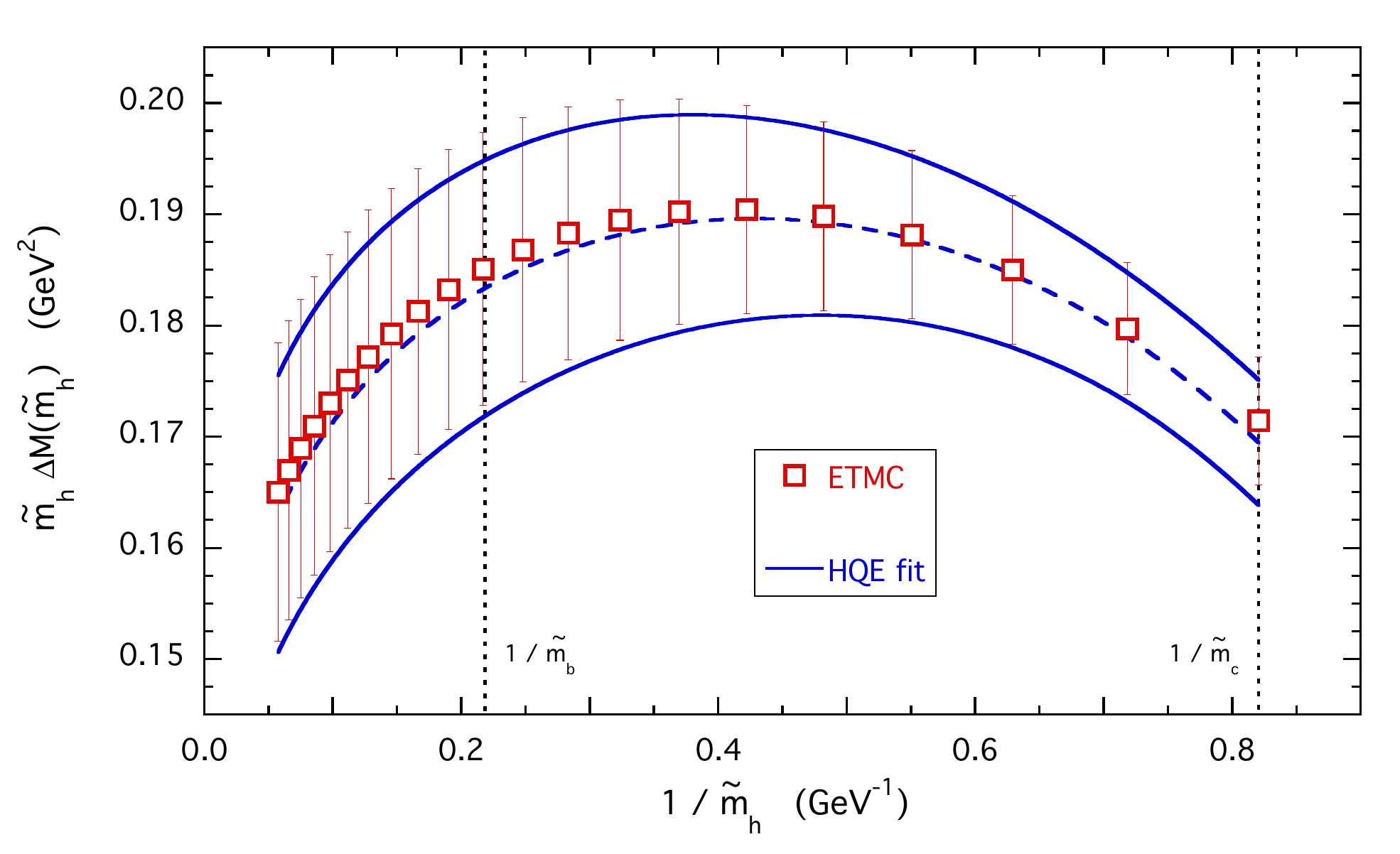}
\vspace{-0.5cm}
\caption{\it \small Lattice data for the quantity $\widetilde{m}_h \Delta M(\widetilde{m}_h)$ (see Eq.~(\ref{eq:DMn})). The dashed and solid lines are the results of the HQE fit (\ref{eq:DM_HQET}), in which the correlation matrix between the lattice data is taken into account. The dashed line corresponds to the central values of the fit, while the solid lines represent one standard deviation. The vertical dotted lines correspond to the positions of the inverse physical $b$-quark and $c$-quark masses, $1 / \widetilde{m}_b$ and $1 / \widetilde{m}_c$, respectively}
\label{fig:DM}
\end{figure}

We have evaluated Eqs.~(\ref{eq:Mavn_final}) and (\ref{eq:DMn}) for $n \lesssim 20$, i.e.~for heavy-quark masses up to $\widetilde{m}_h \simeq 4 \widetilde{m}_b$.
The results are shown in Figs.~\ref{fig:Mav} and \ref{fig:DM}.
It can be seen that, thanks to the definition (\ref{eq:Mavn_final}), the data for the spin-averaged quantity $M_{av}(\widetilde{m}_h) / \widetilde{m}_h$ are quite precise: the uncertainties are at the level of $\simeq 1 \%$ around the charm mass, of $\simeq 0.2 \%$ around the bottom mass and then vanish in the static limit.

Neglecting the effects of dimension-7 operators, the HQE expansion of the heavy-meson masses reads as \cite{Gambino:2012rd}
\bea
     \label{eq:M_HQET}
     \frac{M_{av}(\widetilde{m}_h)}{\widetilde{m}_h} & = & 1 + \frac{\overline{\Lambda}}{\widetilde{m}_h} + 
         \frac{\mu_\pi^2}{2 \widetilde{m}_h^2} + \frac{\rho_D^3 - \rho_{\pi\pi}^3 - \rho_S^3}{4 \widetilde{m}_h^3} ~ , \\[2mm]
     \label{eq:DM_HQET}
     \widetilde{m}_h \Delta M(\widetilde{m}_h) & = & \frac{2}{3} c_G(\widetilde{m}_h, \widetilde{m}_b) \mu_G^2(\widetilde{m}_b) + 
         \frac{\rho_{\pi G}^3 + \rho_A^3 - \rho_{LS}^3}{3 \widetilde{m}_h} ~ , 
\eea
where $\overline{\Lambda}$ is the so-called heavy-quark binding energy, $\mu_\pi^2$ is the matrix element of the kinetic energy operator and the parameters $\rho_i^3$ ($i = D, \pi\pi, S, \pi G, A, LS$) are the matrix elements of the relevant local and non-local operators of dimension-6.
From now on it is understood that all the HQE parameters appearing in Eqs.~(\ref{eq:M_HQET}-\ref{eq:DM_HQET}) are given in the kinetic scheme at the normalization point $\mu_{soft}$, chosen to be equal to $1$ GeV.                                                                       

Taking into account the correlation matrix between the lattice data shown in Figs.~\ref{fig:Mav} and \ref{fig:DM}, the HQE fits (\ref{eq:M_HQET}) and (\ref{eq:DM_HQET}) yield
\bea
    \label{eq:Lambda}
     \overline{\Lambda} & = & 0.551 ~ (13)_{\rm stat} ~ (2)_{\rm syst} ~ \mbox{GeV} = 0.551 ~ (13) ~ \mbox{GeV}~ , \\
    \label{eq:kinetic}
     \mu_\pi^2 & = & 0.314 ~ (14)_{\rm stat} ~ (2)_{\rm syst} ~ \mbox{GeV}^2 = 0.314 ~ (15) ~ \mbox{GeV}^2 ~ , \\
    \label{eq:rho_M}
     \rho_D^3 - \rho_{\pi\pi}^3 - \rho_S^3 & = & 0.174 ~ (12)_{\rm stat} ~ (2)_{\rm syst} ~ \mbox{GeV}^3 = 
                                                                         0.174 ~ (12) ~ \mbox{GeV}^3
 \eea
and
\bea
    \label{eq:chromo}
     \mu_G^2(\widetilde{m}_b) & = & 0.250 ~ (18)_{\rm stat} ~ (8)_{\rm syst} ~ \mbox{GeV}^2 = 0.250 ~ (20) ~ \mbox{GeV}^2 ~ , \\
     \label{eq:rho_DM}
     \rho_{\pi G}^3 + \rho_A^3 - \rho_{LS}^3 & = & - 0.143 ~ (57)_{\rm stat} ~ (21)_{\rm syst} ~ \mbox{GeV}^3 = 
     - 0.143 ~ (60) ~ \mbox{GeV}^3 ~ .
 \eea
The quality of the HQE fits is shown in Figs.~\ref{fig:Mav} and \ref{fig:DM} by the dashed (central values) and solid (one standard deviation) lines.
We stress the remarkable precision obtained for the determinations of $\overline{\Lambda}$ ($\simeq 2.4 \%$), $\mu_\pi^2$ ($\simeq 4.8 \%$), $(\rho_D^3 - \rho_{\pi\pi}^3 - \rho_S^3)$ ($\simeq 6.9 \%$) and $\mu_G^2(\widetilde{m}_b)$ ($\simeq 8.0 \%$), while the quantity  $(\rho_{\pi G}^3 + \rho_A^3 - \rho_{LS}^3)$ has a larger uncertainty ($\simeq 42 \%$).

The HQE fits (\ref{eq:M_HQET}-\ref{eq:DM_HQET}) contain all the terms generated by effective operators up to dimension-6, and in what follows we will refer to the fits (\ref{eq:M_HQET}-\ref{eq:DM_HQET}) as the {\it ``dimension-6''} fit.
We have tried also to include the possible contributions arising from operators of dimension-7, which means that a quartic term has to be added to Eq.~(\ref{eq:M_HQET}) and a quadratic one to Eq.~(\ref{eq:DM_HQET}), viz.
 \bea
       \label{eq:M_HQET_quartic}
       \frac{M_{av}(\widetilde{m}_h)}{\widetilde{m}_h} & = & 1 + \frac{\overline{\Lambda}}{\widetilde{m}_h} + 
            \frac{\mu_\pi^2}{2 \widetilde{m}_h^2} + \frac{\rho_D^3 - \rho_{\pi\pi}^3 - \rho_S^3}{4 \widetilde{m}_h^3} + 
            \frac{\sigma^4}{\widetilde{m}_h^4} ~ , \\[2mm]
     \label{eq:DM_HQET_quadratic}
     \widetilde{m}_h \Delta M(\widetilde{m}_h) & = & \frac{2}{3} c_G(\widetilde{m}_h, \widetilde{m}_b) \mu_G^2(\widetilde{m}_b) + 
         \frac{\rho_{\pi G}^3 + \rho_A^3 - \rho_{LS}^3}{3 \widetilde{m}_h} + \frac{\Delta \sigma^4}{\widetilde{m}_h^2}~ .
 \eea
We obtain
 \bea
    \label{eq:Lambda_quartic}
     \overline{\Lambda} & = & 0.552 ~ (13)_{\rm stat} ~ (2)_{\rm syst} ~ \mbox{GeV} = 0.552 ~ (13) ~ \mbox{GeV}~ , \\
    \label{eq:kinetic_quartic}
     \mu_\pi^2 & = & 0.325 ~ (17)_{\rm stat} ~ (3)_{\rm syst} ~ \mbox{GeV}^2 = 0.325 ~ (17) ~ \mbox{GeV}^2 ~ , \\
    \label{eq:rho_M_quartic}
     \rho_D^3 - \rho_{\pi\pi}^3 - \rho_S^3 & = & 0.133 ~ (34)_{\rm stat} ~ (6)_{\rm syst} ~ \mbox{GeV}^3 = 
                                                                         0.133 ~ (35) ~ \mbox{GeV}^3 ~ , \\
     \label{eq:sigma4_M}
     \sigma^4 & = & 0.0071 ~ (55)_{\rm stat} ~ (10)_{\rm syst} ~ \mbox{GeV}^4 = 0.0071 ~ (55) ~ \mbox{GeV}^4 ~ .  
 \eea
and
\bea
    \label{eq:chromo_quadratic}
     \mu_G^2(\widetilde{m}_b) & = & 0.254 ~ (20)_{\rm stat} ~ (9)_{\rm syst} ~ \mbox{GeV}^2 = 0.254 ~ (22) ~ \mbox{GeV}^2 ~ , \\
     \label{eq:rho_DM_quadratic}
     \rho_{\pi G}^3 + \rho_A^3 - \rho_{LS}^3 & = & - 0.173 ~ (74)_{\rm stat} ~ (25)_{\rm syst} ~ \mbox{GeV}^3 = 
     - 0.173 ~ (79) ~ \mbox{GeV}^3 ~ , \\
      \label{eq:sigma4_DM}
      \Delta \sigma^4 & = & 0.0092 ~ (58)_{\rm stat} ~ (14)_{\rm syst} ~ \mbox{GeV}^4 = 0.0092 ~ (60) ~ \mbox{GeV}^4 ~ . 
 \eea
It can be seen that the values of the HQE parameters related to operators up to dimension-6 are found to be consistent between the {\it ``dimension-6''} and {\it ``dimension-7''} fits.
In particular the results (\ref{eq:Lambda_quartic}), (\ref{eq:kinetic_quartic}) and (\ref{eq:chromo_quadratic}) of the {\it ``dimension-7''} fit confirm nicely both the central values and the uncertainties (\ref{eq:Lambda}), (\ref{eq:kinetic}) and (\ref{eq:chromo}) of the {\it ``dimension-6''} fit.
The result (\ref{eq:rho_DM_quadratic}) is consistent with the corresponding one (\ref{eq:rho_M}) within a larger uncertainty and, finally, the terms (\ref{eq:sigma4_M}) and (\ref{eq:sigma4_DM}) coming from dimension-7 operators are found to be almost consistent with zero.

Note that: 
\begin{itemize}
\item Eqs.~(\ref{eq:kinetic}), (\ref{eq:chromo}) and  Eqs.~(\ref{eq:kinetic_quartic}), (\ref{eq:chromo_quadratic}) imply $(\mu_\pi^2 - \mu_G^2) = 0.064 (19)$ GeV$^2$ for the {\it ``dimension-6''} fit and $(\mu_\pi^2 - \mu_G^2) = 0.072 (22)$ GeV$^2$ for the {\it ``dimension-7''} fit. 
These findings represent a deviation from the so-called BPS limit $\mu_\pi^2 = \mu_G^2$ \cite{Uraltsev:2003ye}.
The deviation is equal to $\approx 20 - 25 \%$ of the kinetic energy term;
\item Eqs.~(\ref{eq:rho_M}), (\ref{eq:rho_DM}) and Eqs.~(\ref{eq:rho_M_quartic}), (\ref{eq:rho_DM_quadratic}) imply $\rho_{\pi \pi}^3 + \rho_S^3 + \rho_{\pi G}^3 + \rho_A^3 = \rho_D^3 + \rho_{LS}^3 - 0.317 (65)$ GeV$^3$ ($0.306 (86)$ GeV$^3$) for the {\it ``dimension-6''}  ({\it ``dimension-7''}) fit.
Since the sum $\rho_{\pi \pi}^3 + \rho_S^3 + \rho_{\pi G}^3 + \rho_A^3$ is always positive definite \cite{Gambino:2012rd}, it follows that $\rho_D^3 + \rho_{LS}^3 \geq 0.317 (65)$ GeV$^3$ ($0.306 (86)$ GeV$^3$) for the {\it ``dimension-6''}  ({\it ``dimension-7''}) fit.
These results show a very sizeable deviation from the BPS limit $\rho_D^3 + \rho_{LS}^3 = 0$ at the level of $\approx 4.9$ (3.6) standard deviations.
\end{itemize}

The correlations among the $b$-quark mass and the HQE parameters of the {\it ``dimension-6''}  and {\it ``dimension-7''} fits are summarized in Tables \ref{tab:correlations_dim6} and \ref{tab:correlations_dim7}, respectively.
The correlations can be taken easily into account by using our bootstrap samplings, which are available upon request.

\begin{table}[htb!]
\begin{center} 
\begin{tabular}{||c||c||c|c|c||c|c||} 
\hline 
& $\widetilde{m}_b$ & $\overline{\Lambda}$ & $\mu_\pi^2$ & $\rho^3$ & $\mu_G^2(\widetilde{m}_b)$ & $\Delta \rho^3$ \\ \hline \hline  
                   $\widetilde{m}_b$ & 1.0 & 0.905 & 0.910 & 0.886 & 0.572 & -0.488 \\ \hline \hline
             $\overline{\Lambda}$ & 0.905 & 1.0 & 0.999 & 0.999 & 0.497 & -0.420 \\ \hline
                           $\mu_\pi^2$ & 0.910 & 0.999 & 1.0 & -0.998 & 0.501 & -0.423 \\ \hline
                                 $\rho^3$ & 0.886 & 0.999 & -0.998 & 1.0 & 0.484 & -0.408 \\ \hline \hline
 $\mu_G^2(\widetilde{m}_b)$ & 0.572 & 0.497 & 0.501 & 0.484 & 1.0 & -0.995 \\ \hline
                      $\Delta \rho^3$ & -0.488 & -0.420 & -0.423 & -0.408 & -0.995 & 1.0 \\ \hline \hline
\end{tabular}
\end{center}
\vspace{-0.25cm}
\caption{\it Correlation matrix among the $b$-quark mass and the HQE parameters of the {\it ``dimension-6''} fit based on Eqs.~(\ref{eq:M_HQET}) and (\ref{eq:DM_HQET}). The quantities $\rho^3$ and $\Delta \rho^3$ stand for $\rho_D^3 - \rho_{\pi\pi}^3 - \rho_S^3$ and $\rho_{\pi G}^3 + \rho_A^3 - \rho_{LS}^3$, respectively.}
\label{tab:correlations_dim6}
\end{table} 

\begin{table}[htb!]
\begin{center} 
\begin{tabular}{||c||c||c|c|c|c||c|c|c||} 
\hline 
& $\widetilde{m}_b$ & $\overline{\Lambda}$ & $\mu_\pi^2$ & $\rho^3$ & $\sigma^4$ & $\mu_G^2(\widetilde{m}_b)$ & $\Delta \rho^3$ & 
   $\Delta \sigma^4$ \\ \hline \hline  
                 $\widetilde{m}_b$ & 1.0 & 0.910 & 0.811 & 0.394 & 0.196 & 0.538 & -0.440 & 0.312 \\ \hline
            $\overline{\Lambda}$ & 0.910 & 1.0 & 0.886 & 0.439 & 0.223 & 0.466 & -0.375 & 0.260 \\ \hline
                          $\mu_\pi^2$ & 0.811 & 0.886 & 1.0 & 0.082 & 0.568 & 0.443 & -0.362 & 0.258 \\ \hline
                                $\rho^3$ & 0.394 & 0.439 & 0.082 & 1.0 & -0.693 & 0.151 & -0.108 & 0.057 \\ \hline
                            $\sigma^4$ & 0.196 & 0.223 & 0.568 & -0.693 & 1.0 & 0.155 & -0.137 & 0.111 \\ \hline \hline
 $\mu_G^2(\widetilde{m}_b)$ & 0.538 & 0.466 & 0.443 & 0.151 & 0.155 & 1.0 & -0.993 & 0.961 \\ \hline
                      $\Delta \rho^3$ & -0.440 & -0.375 & -0.362 & -0.108 & -0.37 & -0.993 & 1.0 & -0.986 \\ \hline
                 $\Delta \sigma^4$ & 0.312 & 0.260 & 0.258 & 0.057 & 0.111 & 0.961 & -0.986 & 1.0 \\ \hline \hline
\end{tabular}
\end{center}
\vspace{-0.25cm}
\caption{\it Correlation matrix among the $b$-quark mass and the HQE parameters of the {\it ``dimension-7''} fit based on Eqs.~(\ref{eq:M_HQET_quartic}) and (\ref{eq:DM_HQET_quadratic}). The quantities $\rho^3$ and $\Delta \rho^3$ stand for $\rho_D^3 - \rho_{\pi\pi}^3 - \rho_S^3$ and $\rho_{\pi G}^3 + \rho_A^3 - \rho_{LS}^3$, respectively.}
\label{tab:correlations_dim7}
\end{table} 

From Table \ref{tab:correlations_dim6} it can be seen that the spin-averaged parameters $\overline{\Lambda}$, $\mu_\pi^2$ and $(\rho_D^3 - \rho_{\pi\pi}^3 - \rho_S^3)$ and, separately, the hyperfine ones $\mu_G^2$ and $(\rho_{\pi G}^3 + \rho_A^3 - \rho_{LS}^3)$ are strongly correlated or anti-correlated among themselves.
Moreover, since our bootstrap sampling takes into account the correlations between the input parameters of Table \ref{tab:8branches}, the data for the spin-averaged meson masses and the hyperfine splitting are partially correlated.
This induces a partial correlation among the hyperfine and the spin-averaged parameters.
Finally, the $b$-quark mass $\widetilde{m}_b$, and correspondingly also the charm mass $\widetilde{m}_c$ [see Eq.~(\ref{eq:correlation_bc})], turn out to be strongly correlated with the spin-averaged HQE parameters and only partially with the hyperfine ones.
In the case of the {\it ``dimension-7''} fit the correlations (see Table \ref{tab:correlations_dim7}) appear to be milder than the corresponding ones of the {\it ``dimension-6''} fit.

As a further consistency check, we have repeated our analysis in the case of the heavy-quark mass dependence of the quantity $M_V^2 - M_{PS}^2$, using the experimental value $M_{D^*}^2 - M_D^2$ at the triggering point.
\begin{figure}[htb!]
\centering{\includegraphics[width=15.5cm]{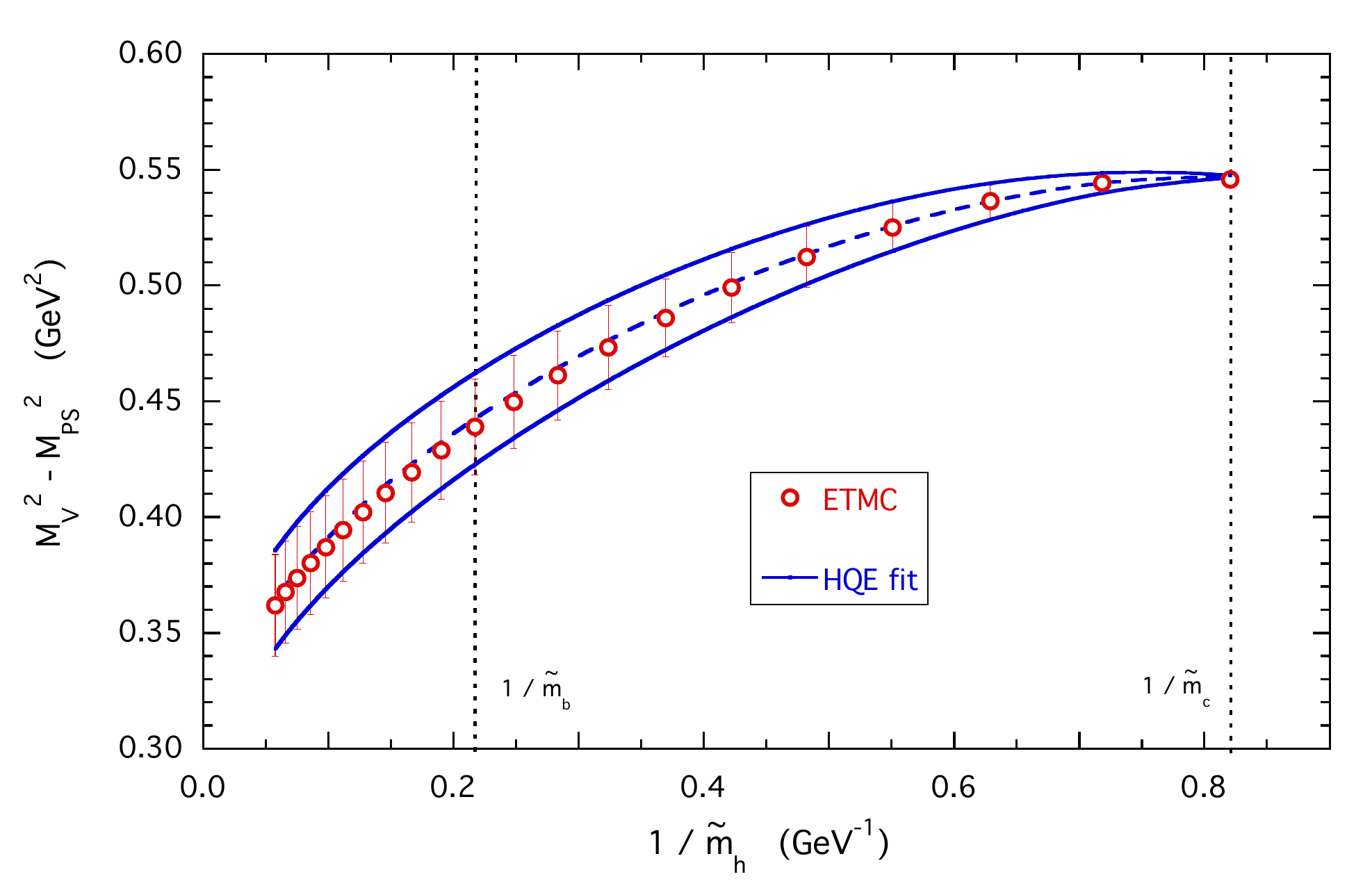}}
\vspace{-0.5cm}
\caption{\it \small Lattice data for the quantity $M_V^2 - M_{PS}^2$ versus the inverse heavy-quark mass $\widetilde{m}_h$. The dashed and solid lines are the result of the HQE fit (\ref{eq:DM2_HQET}), in which  the correlation matrix between the lattice data is taken into account. The dashed line corresponds to the central values of the fits, while the solid lines represent one standard deviation. The vertical dotted lines correspond to the positions of the inverse physical $b$-quark and $c$-quark masses, $1 / \widetilde{m}_b$ and $1 / \widetilde{m}_c$. At the charm mass the experimental value and the error from Ref.~\cite{PDG} are adopted.}
\label{fig:DM2}
\end{figure}
The corresponding data are shown in Fig.~\ref{fig:DM2} and the HQE expansion reads as \cite{Gambino:2012rd}
\be
     M_V^2 - M_{PS}^2 = \frac{4}{3} c_G(\widetilde{m}_h, \widetilde{m}_b) \mu_G^2(\widetilde{m}_b) + \frac{2}{3} 
                                        \frac{\rho_{\pi G}^3 + \rho_A^3 - \rho_{LS}^3 + 2 \overline{\Lambda} \mu_G^2(\widetilde{m}_b)}
                                        {\widetilde{m}_h} + \frac{\Delta \tilde{\rho}^4}{\widetilde{m}_h^2} ~ .
     \label{eq:DM2_HQET}
\ee
Taking into account the correlation matrix between the lattice data, the HQE fit (\ref{eq:DM2_HQET}) (see the solid and dashed lines in Fig.~\ref{fig:DM2}) yields
\bea
    \label{eq:chromo2}
     \mu_G^2(\widetilde{m}_b) & = & 0.270 ~ (17) ~ \mbox{GeV}^2 ~ , \\
     \label{eq:rhop_DM2}
     \rho_{\pi G}^3 + \rho_A^3 - \rho_{LS}^3 + 2 \overline{\Lambda} \mu_G^2(\widetilde{m}_b) & = & 0.164 ~ (46) ~ \mbox{GeV}^3 ~ , \\
     \label{eq:DeltaC}
     \Delta \tilde{\rho}^4 & = & 0.010 ~ (8) ~ \mbox{GeV}^4 ~ .
 \eea
It can be seen that the result (\ref{eq:chromo2}) for $\mu_G^2(\widetilde{m}_b)$ is consistent with the corresponding one given in Eq.~(\ref{eq:chromo_quadratic}).
Using the findings (\ref{eq:Lambda_quartic}) for $\overline{\Lambda}$ and (\ref{eq:chromo2}) for $\mu_G^2(\widetilde{m}_b)$, Eq.~(\ref{eq:rhop_DM2}) implies
 \be
      \rho_{\pi G}^3 + \rho_A^3 - \rho_{LS}^3 = - 0.134 ~ (67) ~ \mbox{GeV}^3 ~ ,
      \label{eq:rho_DM2}
 \ee
which is compatible with the result (\ref{eq:rho_DM_quadratic}) within the uncertainties.

The results (\ref{eq:Lambda}-\ref{eq:rho_DM}) and (\ref{eq:Lambda_quartic}-\ref{eq:sigma4_DM}) of the dimension-6 and dimension-7 fits have been obtained by including radiative corrections up to order $O(\alpha_s^2)$.
Higher order terms might have an impact on the extraction of the HQE parameters, expected to be maximal around the charm mass region.
Therefore, we have applied the dimension-6 fit (\ref{eq:M_HQET}-\ref{eq:DM_HQET}) to the lattice data limiting the range of the heavy-quark masses either to $\widetilde{m}_h \geq 2 \widetilde{m}_c$ or to $\widetilde{m}_h \geq \widetilde{m}_b$.
The corresponding results are shown in Table \ref{tab:comparison_dim6} and compared with the ones obtained in the full range of heavy-quark masses $\widetilde{m}_h \geq \widetilde{m}_c$. 
It can be seen that the parameters $\overline{\Lambda}$, $\mu_\pi^2$ and $\mu_G^2(\widetilde{m}_b)$ (i.e., the matrix elements of operators up to dimension-5) are almost totally insensitive to the range of heavy-quark masses considered, whereas the dimension-6 parameters $\rho_D^3 - \rho_{\pi\pi}^3 - \rho_S^3$ and $\rho_{\pi G}^3 + \rho_A^3 - \rho_{LS}^3$ are only marginally sensitive to the presence of data in the charm region (i.e., consistency within one standard deviation).
 
\begin{table}[htb!]
\begin{center} 
\begin{tabular}{||c||c|c|c||} 
\hline 
 HQE parameter & $\widetilde{m}_h \geq \widetilde{m}_b$ & $\widetilde{m}_h \geq 2 \widetilde{m}_c$ & $\widetilde{m}_h \geq \widetilde{m}_c$ 
 \\ \hline \hline
                                        $\overline{\Lambda}$ (GeV) &  ~ 0.552 ~ (13) & ~ 0.552 ~ (13) & ~ 0.551 ~ (13) \\ \hline
                                               $\mu_\pi^2$ (GeV$^2$) &  ~ 0.325 ~ (15) & ~ 0.323 ~ (16) & ~ 0.314 ~ (15) \\ \hline
      $\rho_D^3 - \rho_{\pi\pi}^3 - \rho_S^3$ (GeV$^3$) &  ~ 0.146 ~ (31) & ~ 0.153 ~ (24) & ~ 0.174 ~ (12) \\ \hline \hline
                      $\mu_G^2(\widetilde{m}_b)$ (GeV$^2$) & ~ 0.253 ~ (22) & ~ 0.254 ~ (22) & ~ 0.250 ~ (20) \\ \hline
 $\rho_{\pi G}^3 + \rho_A^3 - \rho_{LS}^3$ (GeV$^3$) & -0.133 ~ (69) & -0.158 ~ (70) & -0.143 ~ (60) \\ \hline \hline
\end{tabular}
\end{center}
\vspace{-0.25cm}
\caption{\it Results obtained for the HQE parameters $\overline{\Lambda}$, $\mu_\pi^2$ , $\rho_D^3 - \rho_{\pi\pi}^3 - \rho_S^3$, $\mu_G^2(\widetilde{m}_b)$ and $\rho_{\pi G}^3 + \rho_A^3 - \rho_{LS}^3$ for different ranges of the heavy-quark mass $\widetilde{m}_h$ included in the {\it ``dimension-6''} fit (\ref{eq:M_HQET}-\ref{eq:DM_HQET}).}
\label{tab:comparison_dim6}
\end{table} 

In order to obtain our final determinations of the HQE parameters we perform the average of the results corresponding to the {\it ``dimension-6''} and {\it ``dimension-7''} fits as well as to the {\it ``dimension-6''} fit with the range of the heavy-quark masses limited to $\widetilde{m}_h \geq 2 \widetilde{m}_c$ (see third column of Table \ref{tab:comparison_dim6}). 
The average and the corresponding uncertainty are evaluated according to Eq.~(28) of Ref.~\cite{Carrasco:2014cwa}.
Moreover, we want to consider the impact of the uncertainty in the conversion from the $\overline{MS}$ scheme to the kinetic one at the charm mass on the extracted HQE parameters
As a matter of fact, a systematic shift of the value of $\widetilde{m}_c$ can propagate into the chain of the heavy-quark masses leading to a change of the values of the extracted HQE parameters.
Thus, we have shifted the value of $\widetilde{m}_c$ by 40 MeV\footnote{In Ref.~\cite{Gambino:2011cq} the uncertainty in the conversion from $\overline{m}_c(\overline{m}_c)$ to $\widetilde{m}_c$ has been estimated to be $\approx 20$ MeV. We have conservatively doubled that uncertainty.} and repeated our whole analysis, obtaining a change equal to 0.150 GeV for $\widetilde{m}_b$, 0.022 GeV for $\overline{\Lambda}$, 0.027 GeV$^2$ for $\mu_\pi^2$, $0.017$ GeV$^3$ for $\rho_D^3 - \rho_{\pi \pi}^3 - \rho_S^3$, 0.013 GeV$^2$ for $\mu_G^2(m_b)$ and 0.045 GeV$^3$ for $\rho_{\pi G}^3 + \rho_A^3 - \rho_{LS}^3$.

The inclusion of the above uncertainties (added in quadrature) lead to the final results
 \bea
      \widetilde{m}_c & = & 1.219 ~ (41) ~ (40)_{conv} ~ \mbox{GeV} = 1.219 ~ (57) ~ \mbox{GeV} ~ , \\
      \widetilde{m}_b & = & 4.605 ~ (132) ~ (150)_{conv} ~ \mbox{GeV} = 4.605 ~ (201) ~ \mbox{GeV} ~ , \\
      \nonumber \\
       \overline{\Lambda} & = & 0.552 ~ (13) ~ (22)_{conv} ~ \mbox{GeV} = 0.552 ~ (26) ~ \mbox{GeV} ~ , \\
       \label{eq:mupi2}
        \mu_\pi^2 & = & 0.321 ~ (17) ~ (27)_{conv} ~ \mbox{GeV}^2 =  0.321 ~ (32) ~ \mbox{GeV}^2 ~ , \\
        \rho_D^3 - \rho_{\pi \pi}^3 - \rho_S^3 & = & 0.153 ~ (30) ~ (17)_{conv} ~ \mbox{GeV}^3 =  0.153 ~ (34) ~ \mbox{GeV}^3 ~ , \\
        \nonumber \\
        \mu_G^2(m_b) & = & 0.253 ~ (21) ~ (13)_{conv} ~ \mbox{GeV}^2 = 0.253 ~ (25) ~ \mbox{GeV}^2 ~ ,  \\      
        \rho_{\pi G}^3 + \rho_A^3 - \rho_{LS}^3 & = & -0.158 ~ (71) ~ (45)_{conv} ~ \mbox{GeV}^3 = -0.158 ~ (84) ~ \mbox{GeV}^3 ~ ,
 \eea
where $()_{conv}$ indicates the errors generated by the uncertainty in the conversion from the $\overline{MS}$ scheme to the kinetic one at the charm mass.
The reduction of this source of uncertainty will certainly deserve future investigations. 

Before closing this Section, we want to comment briefly on the relation between our results and those obtained in recent analyses of the inclusive semileptonic B-meson decays~\cite{Gambino:2016jkc,Alberti:2014yda}.

We start by warning the reader that in this work $\mu_\pi^2$ and $\mu^2_G(m_b)$ refer to asymptotic matrix elements, i.e.~matrix elements of asymptotically heavy mesons, while the inclusive semileptonic fits are sensitive to the matrix elements of the same operators in the physical B-meson.
The relations between the two concepts are
\bea
    \mu_\pi^2|_B & = & \mu_\pi^2|_\infty - \frac{\rho_{\pi\pi}^3 + \frac{1}{2} \rho_{\pi G}^3}{\widetilde{m}_b} + {\cal{O}}(1 / \widetilde{m}_b^2) ~ , \\
    \mu_G^2(m_b)|_B & = & \mu_G^2(m_b)|_\infty + \frac{\rho_S^3 + \rho_A^3 + \frac{1}{2} \rho_{\pi G}^3}{\widetilde{m}_b} + 
                                            {\cal{O}}(1 / \widetilde{m}_b^2) ~ .
\eea
It should also be kept in mind that the semileptonic fits are not very sensitive to $\mu_G^2(m_b)$ and $\rho_{LS}^3$, which are mostly determined by loose constraints based on heavy quark sum rules. 
In particular the constraint $\mu_G^2(m_b)|_B = 0.35(7)$ GeV$^2$ is applied in Refs.~\cite{Gambino:2016jkc,Alberti:2014yda}. 
As a first application of our results we can check their consistency with this constraint.
The values $\mu_\pi^2|_B = 0.432 (68)$ GeV$^2$ and $\mu_\pi^2|_B = 0.465 (68)$ GeV$^2$ were found in Refs.~\cite{Gambino:2016jkc,Alberti:2014yda}, respectively, which differ only for the inclusion of higher-order power corrections. 
Comparing these values with our final result (\ref{eq:mupi2}), it follows that the combination $\rho_{\pi\pi}^3 + \frac{1}{2} \rho_{\pi G}^3$ should be large and negative, $-0.51 (35)$ GeV$^2$, where we have taken the smaller value of $\mu_\pi^2|_B$ from Ref.~\cite{Gambino:2016jkc}.
Since the sum $\rho_{\pi\pi}^3 + \rho_S^3 + \rho_A^3 + \rho_{\pi G}^3$ is positive by definition, it also follows that $\rho_S^3 + \rho_A^3 + \frac{1}{2} \rho_{\pi G}^3 > 0.51 (35)$ GeV$^2$, or $\mu_G^2|_B > \mu^2_G|_\infty + 0.11 (8)$ GeV$^2$ $= 0.36 (8)$ GeV$^2$.
Despite the large errors, there is a clear indication that the constraint employed in the semileptonic fits is adequate.
We also note that the large values taken by some of the non-local matrix elements are consistent with the observations made in Ref.~\cite{Gambino:2012rd}. 
A detailed discussion of our results in the context of the heavy quark sum rules and in particular of  the zero recoil sum rule is postponed to a future publication.

Of course, in order to employ our results in other observables, like the inclusive semileptonic decay rates of the B-meson, it is necessary that all the matrix elements are defined as short distance quantities, not affected by renormalons.
As is well known, the OPE of the inclusive semileptonic $B$-meson decay rate predicts \cite{OPE} that the corrections to the free-quark decay rate are suppressed by two powers of the $b$-quark mass and can be parameterized in terms of the HQE matrix elements $\mu_\pi^2$ and $\mu_G^2(m_b)$.
In terms of the heavy-quark pole mass the radiative corrections to the free-quark decay rate are plagued by renormalons, which however are cancelled out when the pole mass is replaced in favor of a short-distance heavy-quark mass \cite{Beneke:1994sw,Beneke:1994bc}.
This is a crucial feature for the OPE analysis of the inclusive semileptonic $B$-meson decays, since the appearance of renormalons in the radiative corrections of the leading-order decay rate may signal the presence of non-perturbative corrections in the inverse heavy-quark mass, that cannot be parameterized using the same HQE matrix elements $\mu_\pi^2$ and $\mu_G^2(m_b)$ extracted from the analysis of heavy-meson masses.
In principle, the kinetic scheme is designed to achieve precisely that.

\section{Conclusions}
\label{sec:conclusions}

We have presented a precise lattice computation of pseudoscalar and vector heavy-light meson masses for heavy-quark masses ranging from the physical charm mass up to $\simeq 4$ times the physical b-quark mass, adopting the gauge configurations generated by the European Twisted Mass Collaboration (ETMC) with $N_f = 2+1+1$ dynamical quarks at three values of the lattice spacing ($a \simeq 0.062, 0.082, 0.089$ fm) with pion masses in the range $M_\pi \simeq 210 - 450$ MeV. 
The heavy-quark mass has been simulated directly on the lattice up to $\simeq 3$ times the physical charm mass. 
The interpolation to the physical $b$-quark mass has been performed using the {\it ETMC ratio} method, based on ratios of the spin-averaged meson masses computed at nearby heavy-quark masses.

The kinetic mass scheme has been adopted in order to work with a short-distance mass free from renormalon ambiguites (also often used in the analysis of the inclusive semileptonic $B$-meson decays relevant for the determination of the CKM entry $V_{cb}$).
The extrapolation to the physical pion mass and to the continuum limit yields $m_b^{\rm kin}(1~\mbox{GeV}) = 4.61 (20)$ GeV, which corresponds to $\overline{m}_b(\overline{m}_b) = 4.26 (18)$ GeV in the $\overline{MS}$ scheme, and is in agreement with the results of the OPE analysis of the inclusive semileptonic $B$-meson decays \cite{Gambino:2016jkc,Alberti:2014yda}.

Then the {\it ratio} method has been applied above the physical $b$-quark mass to provide heavy-light meson masses towards the static point. 
The lattice data have been analyzed in terms of the Heavy Quark Expansion and the matrix elements of dimension-4 and dimension-5 operators have been determined with a good precision, namely:
 \bea
        \label{eq:dim4_final}
        \overline{\Lambda} & = & 0.552 ~ (26) ~\mbox{GeV} ~ , \\
        \label{eq:dim5_1_final}
        \mu_\pi^2 & = & 0.321 ~ (32)~\mbox{GeV}^2 ~ , \\
         \label{eq:dim5_2_final}
        \mu_G^2(m_b) & = & 0.253 ~ (25)~\mbox{GeV}^2 ~ .
  \eea
The data has allowed also to estimate the size of two combinations of the matrix elements of dimension-6 operators, namely: 
 \bea
     \label{eq:dim6_1_final}
     \rho_D^3 - \rho_{\pi \pi}^3 - \rho_S^3 & = & 0.153 ~ (34) ~\mbox{GeV}^3  ~ , \\
     \label{eq:dim6_2_final}
     \rho_{\pi G}^3 + \rho_A^3 - \rho_{LS}^3 & = & -0.158 ~ (84) ~\mbox{GeV}^3 ~ .
 \eea
All the above HQE parameters, as well as the physical $c$- and $b$-quark masses, are found to be highly correlated and therefore the full covariance matrix has been provided (see Tables \ref{tab:correlations_dim6}-\ref{tab:correlations_dim7}).
We stress that our results (\ref{eq:dim4_final}-\ref{eq:dim6_2_final}), which are specific to the kinetic scheme, represent the first unquenched lattice determinations of the HQE parameters.

The extracted dimension-5 and dimension-6 HQE parameters play a crucial role in the OPE analysis of the inclusive semileptonic $B$-meson decays relevant for the determination of the CKM entries $V_{ub}$ and $V_{cb}$. 
Our findings may help validating and possibly improving the inclusive determination of these fundamental parameters of the Standard Model.

\section*{Acknowledgements}
We warmly thank P.~Dimopoulos, R.~Frezzotti, V.~Lubicz, G.~Martinelli and C.~Tarantino for fruitful discussions and their continuous support.
We thank the ETMC members for having generated and made publicly available the gauge configurations used for this study. 
We gratefully acknowledge the CPU time provided by PRACE under the project PRA067 {\it ``First Lattice QCD study of B-physics with four flavors of dynamical quarks"} and by CINECA under the specific initiative INFN-LQCD123 on the BG/Q system Fermi at CINECA (Italy).


\begin{thebibliography}{99}

\bibitem{Gambino:2016jkc}
  P.~Gambino, K.~J.~Healey and S.~Turczyk, 
  Phys.\ Lett.\ B {\bf 763} (2016) 60 
  [arXiv:1606.06174 [hep-ph]].

\bibitem{Alberti:2014yda} 
  A.~Alberti, P.~Gambino, K.~J.~Healey and S.~Nandi, 
  Phys.\ Rev.\ Lett.\  {\bf 114} (2015) no.6,  061802 
  [arXiv:1411.6560 [hep-ph]].  

\bibitem{PDG}
  C.~Patrignani {\it et al.} [Particle Data Group],
   Chin.\ Phys.\ C {\bf 40} (2016) no.10,  100001.

  \bibitem{Bigi:2017njr}
  D.~Bigi, P.~Gambino and S.~Schacht,
  arXiv:1703.06124 [hep-ph].

\bibitem{Gambino:2012rd}
  P.~Gambino, T.~Mannel and N.~Uraltsev,
  JHEP {\bf 1210} (2012) 169
  [arXiv:1206.2296 [hep-ph]].

\bibitem{Blossier:2009hg}
  B.~Blossier {\it et al.} [ETM Coll.],
  JHEP {\bf 1004} (2010) 049
  [arXiv:0909.3187 [hep-lat]].

\bibitem{Dimopoulos:2011gx}
  P.~Dimopoulos {\it et al.} [ETM Coll.],
  JHEP {\bf 1201} (2012) 046
  [arXiv:1107.1441 [hep-lat]].

\bibitem{Carrasco:2013zta}
  N.~Carrasco {\it et al.} [ETM Coll.],
  JHEP {\bf 1403} (2014) 016
  [arXiv:1308.1851 [hep-lat]].

\bibitem{Bussone:2016iua}
  A.~Bussone {\it et al.} [ETM Coll.],
  Phys.\ Rev.\ D {\bf 93} (2016) no.11,  114505
  [arXiv:1603.04306 [hep-lat]].

\bibitem{Bigi:1994em}
  I.~I.~Y.~Bigi, M.~A.~Shifman, N.~G.~Uraltsev and A.~I.~Vainshtein,
  Phys.\ Rev.\ D {\bf 50} (1994) 2234
  [hep-ph/9402360].

\bibitem{Bigi:1996si}
  I.~I.~Y.~Bigi, M.~A.~Shifman, N.~Uraltsev and A.~I.~Vainshtein,
  Phys.\ Rev.\ D {\bf 56} (1997) 4017
  [hep-ph/9704245].

\bibitem{Beneke:1994sw}
  M.~Beneke and V.~M.~Braun,
  Nucl.\ Phys.\ B {\bf 426} (1994) 301
  [hep-ph/9402364].

\bibitem{Luke:1994xd}
  M.~E.~Luke, A.~V.~Manohar and M.~J.~Savage,
  Phys.\ Rev.\ D {\bf 51} (1995) 4924
  [hep-ph/9407407].

\bibitem{Martinelli:1995vj}
  G.~Martinelli and C.~T.~Sachrajda,
  Phys.\ Lett.\ B {\bf 354} (1995) 423
  [hep-ph/9502352].

\bibitem{Gambino:2011cq}
  P.~Gambino,
  JHEP {\bf 1109} (2011) 055
  [arXiv:1107.3100 [hep-ph]].

\bibitem{FLAG}
  S.~Aoki {\it et al.},
  Eur.\ Phys.\ J.\ C {\bf 77} (2017) no.2,  112
  [arXiv:1607.00299 [hep-lat]].

\bibitem{Heitger:2004gb}
  J.~Heitger {\it et al.} [ALPHA Coll.],
  JHEP {\bf 0411} (2004) 048
  [hep-ph/0407227].

\bibitem{Guazzini:2007bu}
  D.~Guazzini {\it et al.} [ALPHA Coll.],
  JHEP {\bf 0710} (2007) 081
  [arXiv:0705.1809 [hep-lat]].

\bibitem{Grozin:2007fh}
  A.~G.~Grozin, P.~Marquard, J.~H.~Piclum and M.~Steinhauser,
  Nucl.\ Phys.\ B {\bf 789} (2008) 277
  [arXiv:0707.1388 [hep-ph]].

\bibitem{Ewing:1995ih}
  A.~K.~Ewing {\it et al.} [UKQCD Collaboration],
  Phys.\ Rev.\ D {\bf 54} (1996) 3526
  [hep-lat/9508030].

 \bibitem{Gimenez:1996nw}
  V.~Gimenez, G.~Martinelli and C.~T.~Sachrajda,
  Phys.\ Lett.\ B {\bf 393} (1997) 124
  [hep-lat/9607018].

  \bibitem{Gimenez:1996av}
  V.~Gimenez, G.~Martinelli and C.~T.~Sachrajda,
  Nucl.\ Phys.\ B {\bf 486} (1997) 227
  [hep-lat/9607055].

\bibitem{Kronfeld:2000gk}
  A.~S.~Kronfeld and J.~N.~Simone,
  Phys.\ Lett.\ B {\bf 490} (2000) 228
  Erratum: [Phys.\ Lett.\ B {\bf 495} (2000) 441]
  [hep-ph/0006345].

\bibitem{Komijani:2016jrh}
  J.~Komijani {\it et al.},
  arXiv:1611.07411 [hep-lat].

\bibitem{Baron:2010bv}
  R.~Baron {\it et al.} [ETM Coll.],
  JHEP {\bf 1006} (2010) 111
  [arXiv:1004.5284 [hep-lat]].
   
 \bibitem{Baron:2011sf}
  R.~Baron {\it et al.}  [ETM Coll.],
  PoS LATTICE {\bf 2010} (2010) 123
  [arXiv:1101.0518 [hep-lat]].

\bibitem{Carrasco:2014cwa}
  N.~Carrasco {\it et al.} [ETM Coll.],
  Nucl.\ Phys.\ B {\bf 887} (2014) 19
  [arXiv:1403.4504 [hep-lat]].

\bibitem{Iwasaki:1985we}
  Y.~Iwasaki,
  Nucl.\ Phys.\ B {\bf 258} (1985) 141.

\bibitem{Frezzotti:2000nk}
 R.~Frezzotti {\it et al.} [Alpha Coll.],
 JHEP {\bf 0108} (2001) 058
 [hep-lat/0101001].

\bibitem{Frezzotti:2003xj}
  R.~Frezzotti and G.C.~Rossi,
  Nucl.\ Phys.\ Proc.\ Suppl.\  {\bf 128} (2004) 193
  [hep-lat/0311008].
  
\bibitem{Frezzotti:2003ni}
  R.~Frezzotti and G.C.~Rossi,
  JHEP {\bf 0408} (2004) 007
  [hep-lat/0306014].

\bibitem{Osterwalder:1977pc}
  K.~Osterwalder and E.~Seiler,
  Annals Phys.\  {\bf 110} (1978) 440.

\bibitem{Frezzotti:2004wz}
  R.~Frezzotti and G.C.~Rossi,
  JHEP {\bf 0410} (2004) 070
  [hep-lat/0407002].

\bibitem{Foster:1998vw}
  M.~Foster {\it et al.} [UKQCD Coll.],
  Phys.\ Rev.\ D {\bf 59} (1999) 074503
  [hep-lat/9810021].

\bibitem{McNeile:2006bz}
  C.~McNeile {\it et al.} [UKQCD Coll.],
  Phys.\ Rev.\ D {\bf 73} (2006) 074506
  [hep-lat/0603007].

\bibitem{Gusken:1989qx}
  S.~Gusken,
  Nucl.\ Phys.\ Proc.\ Suppl.\  {\bf 17} (1990) 361.

\bibitem{Albanese:1987ds}
  M.~Albanese {\it et al.} [APE Coll.],
  Phys.\ Lett.\ B {\bf 192} (1987) 163.

\bibitem{Blossier:2009kd}
  B.~Blossier, M.~Della Morte, G.~von Hippel, T.~Mendes and R.~Sommer,
  JHEP {\bf 0904} (2009) 094
  [arXiv:0902.1265 [hep-lat]].

\bibitem{Czarnecki:1997sz}
  A.~Czarnecki, K.~Melnikov and N.~Uraltsev,
  Phys.\ Rev.\ Lett.\  {\bf 80} (1998) 3189
  [hep-ph/9708372].

\bibitem{Chetyrkin:1999pq}
  K.~G.~Chetyrkin and A.~Retey,
  Nucl.\ Phys.\ B {\bf 583} (2000) 3
  [hep-ph/9910332].

\bibitem{Lubicz:2016bbi}
  V.~Lubicz, A.~Melis and S.~Simula,
  arXiv:1610.09671 [hep-lat].

\bibitem{Uraltsev:2003ye}
  N.~Uraltsev,
  Phys.\ Lett.\ B {\bf 585} (2004) 253
  [hep-ph/0312001].

\bibitem{OPE}

  J.~Chay, H.~Georgi and B.~Grinstein,
  Phys.\ Lett.\ B {\bf 247} (1990) 399.

  I.~I.~Y.~Bigi, N.~G.~Uraltsev and A.~I.~Vainshtein,
  Phys.\ Lett.\ B {\bf 293} (1992) 430
   Erratum: [Phys.\ Lett.\ B {\bf 297} (1992) 477]
  [hep-ph/9207214].
  
  I.~I.~Y.~Bigi, M.~A.~Shifman, N.~G.~Uraltsev and A.~I.~Vainshtein,
  Phys.\ Rev.\ Lett.\  {\bf 71} (1993) 496
  [hep-ph/9304225].

\bibitem{Beneke:1994bc}
  M.~Beneke, V.~M.~Braun and V.~I.~Zakharov,
  Phys.\ Rev.\ Lett.\  {\bf 73} (1994) 3058
  [hep-ph/9405304].
  
\end{thebibliography}
\end{document}